\documentclass[journal]{IEEEtran}

\usepackage{algorithm}
\usepackage[noend]{algpseudocode}

\usepackage{amsmath}
\usepackage{mathtools}
\usepackage{amsmath,amsfonts,amssymb}
\usepackage{booktabs}
\usepackage{mathrsfs}
\usepackage{makecell}
\usepackage{multirow}
\usepackage{xcolor}
\usepackage[noadjust]{cite}

\usepackage[switch,columnwise]{lineno}
\usepackage{amssymb}
\usepackage{pifont}
\usepackage{upgreek}

\begin{document}
\bstctlcite{IEEEexample:BSTcontrol}
\title{NeuralTree: A 256-Channel 0.227$\upmu$J/class Versatile Neural Activity Classification and Closed-Loop Neuromodulation SoC}

\author{Uisub~Shin,~\IEEEmembership{Student Member,~IEEE,}~Cong~Ding,~\IEEEmembership{Student Member,~IEEE,}~Bingzhao~Zhu,~\IEEEmembership{Student Member,~IEEE,}\\~Yashwanth~Vyza,~Alix~Trouillet,~Emilie~C.~M.~Revol,~\IEEEmembership{Student Member,~IEEE,}~St\'ephanie~P.~Lacour,\\~\IEEEmembership{Member,~IEEE,} and~Mahsa~Shoaran,~\IEEEmembership{Member,~IEEE\vspace{-6mm}}%

\thanks{Manuscript received May 6, 2022}%
\thanks{Uisub Shin is with the Institute of Electrical and Micro Engineering, EPFL, 1202 Geneva, Switzerland, and the School of Electrical and Computer Engineering, Cornell University, Ithaca, NY 14853, USA (e-mail: us52@cornell.edu).}
\thanks{Cong Ding, Yashwanth Vyza, Alix Trouillet, Emilie C. M. Revol,  St\'ephanie P. Lacour, and Mahsa Shoaran are  with the Institute of Electrical and Micro Engineering and Center for Neuroprosthetics, EPFL, 1202 Geneva, Switzerland.}
\thanks{Bingzhao Zhu is with the Institute of Electrical and Micro Engineering, EPFL, 1202 Geneva, Switzerland, and the School of Applied and Engineering Physics, Cornell University, Ithaca, NY 14853, USA.}
}

\markboth{$>$ REPLACE THIS LINE WITH YOUR MANUSCRIPT ID NUMBER (DOUBLE-CLICK HERE TO EDIT) $<$}%
{Shell \MakeLowercase{\textit{et al.}}: Bare Demo of IEEEtran.cls for IEEE Journals}

\maketitle

\begin{abstract}
Closed-loop neural interfaces with on-chip machine learning can detect and suppress disease symptoms in neurological disorders or restore lost functions in paralyzed patients. While high-density neural recording can provide rich neural activity information for accurate disease-state detection, existing systems have low channel count and poor scalability, which could limit their therapeutic efficacy. This work presents a highly scalable and versatile closed-loop neural interface SoC that can overcome these limitations. A 256-channel time-division multiplexed (TDM) front-end with a two-step fast-settling mixed-signal DC servo loop (DSL) is proposed to record high-spatial-resolution neural activity and perform channel-selective brain-state inference. A tree-structured neural network (NeuralTree) classification processor extracts a rich set of neural biomarkers in a patient- and disease-specific manner. Trained with an energy-aware learning algorithm, the NeuralTree classifier detects the  symptoms of  underlying disorders (e.g., epilepsy and movement disorders) at an optimal energy-accuracy trade-off. A 16-channel high-voltage (HV) compliant neurostimulator closes the therapeutic loop by delivering charge-balanced biphasic current pulses to the brain. The proposed SoC was fabricated in 65nm CMOS and achieved a 0.227$\boldsymbol\upmu$J/class energy efficiency in a compact area of 0.014mm\textsuperscript{2}/channel. The SoC was extensively verified on human electroencephalography (EEG) and intracranial EEG (iEEG) epilepsy datasets, obtaining 95.6\%/94\% sensitivity and 96.8\%/96.9\% specificity, respectively. \emph{In-vivo} neural recordings using soft $\boldsymbol\upmu$ECoG arrays and multi-domain biomarker extraction were further performed on a rat model of epilepsy. In addition, for the first time in literature, on-chip classification of rest-state tremor in Parkinson's disease from human local field potentials (LFPs) was demonstrated. 
\end{abstract}

\begin{IEEEkeywords}
machine learning, neural network, decision tree, closed-loop neuromodulation, epilepsy, Parkinson's disease, energy-efficient classification, seizure, tremor 
\end{IEEEkeywords}

\section{Introduction}
\IEEEPARstart{N}{eurological} disorders are the second leading cause of global deaths and the leading cause of disability worldwide \cite{feigin2017global}. Epilepsy ($>$60M) and Parkinson's disease ($>$10M) are among common examples, and many patients are living with medication-refractory symptoms of these disorders. Closed-loop brain stimulation has emerged as a promising therapeutic solution to treating such disorders \cite{skarpaas2009intracranial, meidahl2017adaptive, zhu2021closed, altaf201516, shoaran201616}, and a number of FDA-approved and research-based devices are currently available, including NeuroPace's responsive neurostimulation \cite{jarosiewicz2021rns} and Medtronic's Percept deep-brain stimulation (DBS) \cite{jimenez2021device} systems. However, these devices have a low channel count (4\text{--}6) and rely on  simplistic symptom detection algorithms (e.g., feature thresholding), which could result in suboptimal detection accuracy and limited therapeutic efficacy. 

Over the past decade, we have witnessed a growing adoption of machine learning (ML) for symptom prediction in a variety of neurological  conditions such as epilepsy \cite{shoeb2009application}, Parkinson's disease (PD) \cite{yao2020improved}, depression \cite{sani2018mood}, and memory disorders~\cite{ezzyat2018closed}. Through neural signal acquisition, biomarker extraction, and ML-based classification, pathological disease states can be detected more accurately and suppressed more effectively than conventional methods. Moreover, ML intelligence can improve the accuracy of motor intention decoding in brain-machine interfaces (BMIs) for rehabilitation of motor impairments \cite{xie2018decoding, yao2022fast}. Despite recent innovations, existing ML-embedded SoCs \cite{altaf201516, cheng2018fully, shoaran2018energy, o2018recursive, o202026, huang20191, wang2021closed, chua20211, zhang2022patient} are limited in the following aspects: 

\subsubsection{Low channel count}
\indent As the analog front-end (AFE) often dominates the overall area of a neural interface system, the channel count of existing ML-SoCs is limited to  8--32, which may not be sufficient to collect clinically meaningful information for accurate disease state prediction.

\subsubsection{Low hardware efficiency}
\indent Despite low channel count, the area and energy consumption of the existing SoCs are prohibitive, making it difficult to scale up the number of channels. This is mainly due to their hardware complexity that grows proportional to the number of channels and biomarkers. 

\subsubsection{Limited application}
\indent With limited sets of biomarkers and conventional classifiers, most neural interface SoCs reported so far have targeted a single task, epileptic seizure detection, while there exist many other conditions that could benefit from the closed-loop neural interface technology.

\begin{figure}[t]
  \centering
  \includegraphics[width=1\columnwidth]{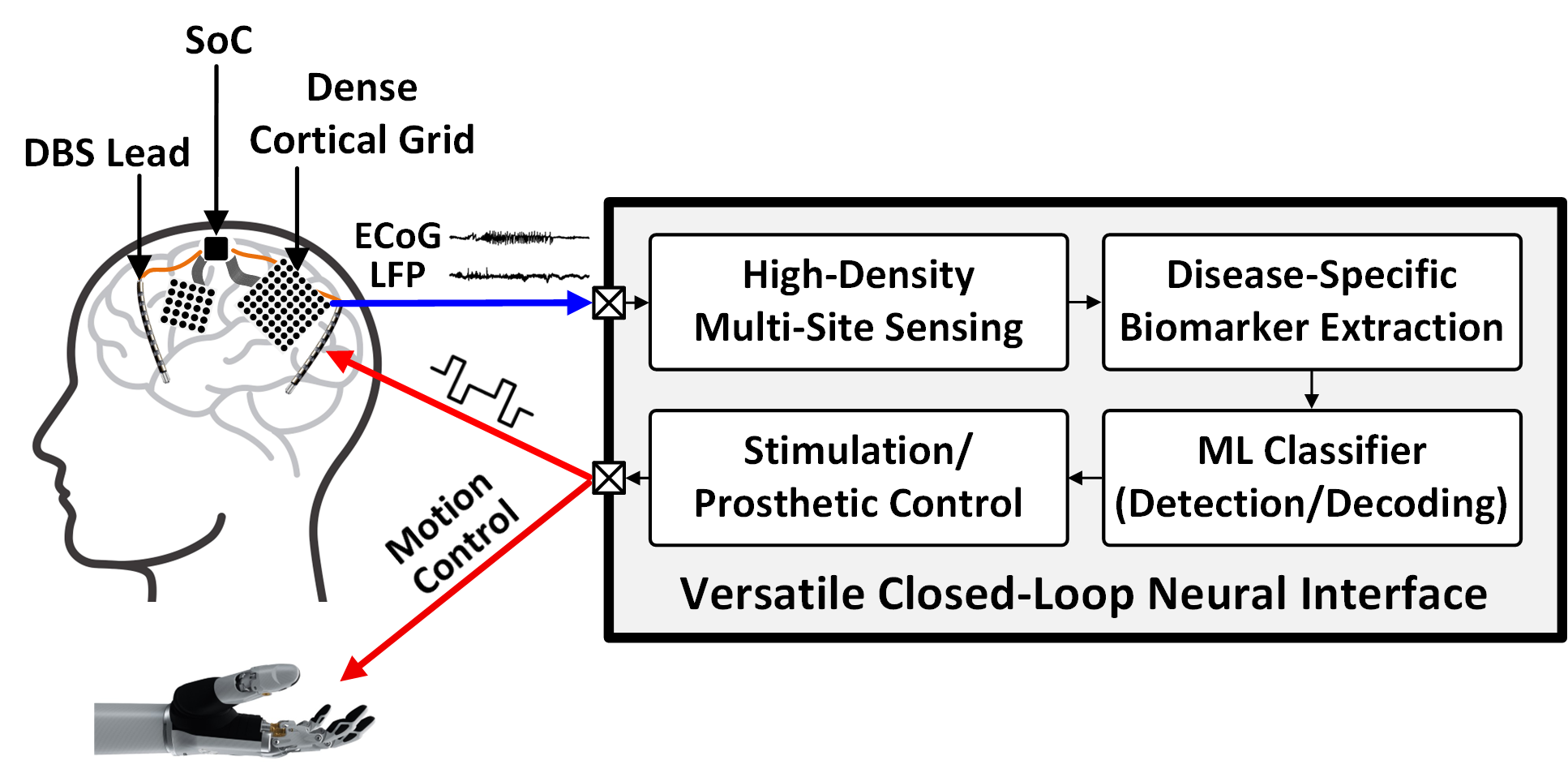}
  \vspace{-3mm}
  \caption{Versatile closed-loop neural interface platform with high-density sensing and stimulation capabilities.}\vspace{-3mm}
  \label{f1_nextgen}
\end{figure}

Closed-loop SoCs would advance further with a higher number of channels and greater adaptability to various neural classification tasks. In epileptic seizure detection, for instance, covering a larger brain area with a high-spatial-resolution electrocorticography (ECoG) array will enable more precise localization of epileptic foci and better mapping of seizure onset \cite{chang2015towards}, thus enhancing the seizure detection accuracy of the trained classifier. In treating movement disorders such as PD and essential tremor, high-density DBS can engage target brain regions more effectively while reducing stimulation-induced side effects \cite {anderson2018optimized}. Furthermore,  high channel count can enhance the accuracy of motor intention decoding in prosthetic BMIs by collecting higher-resolution motor and sensory information~\cite{kaiju2017high}. Fig. \ref{f1_nextgen} presents the envisioned versatile neural interface platform that integrates a large number of channels. A patient- and disease-specific classifier detects the pathological symptoms of brain disorders and activates a neurostimulator to provide cortical or deep-brain stimulation for symptom suppression. In BMIs, motor intention can be decoded to control prosthetic devices and provide sensory feedback to the brain via electrical stimulation. This high-density, versatile neural interface device could advance our understanding of  complex brain dynamics and provide new therapeutic opportunities for people suffering from various neurological/psychiatric disorders and motor conditions. A miniaturized, energy-efficient implementation of such devices is a key to enabling  next-generation closed-loop neural SoCs.     

To overcome the aforementioned limitations of existing systems and transition towards the next-generation closed-loop neural interface, this paper presents a 256-channel highly scalable, energy-efficient neural activity classification and closed-loop neuromodulation SoC. A 256-channel area-efficient AFE collects high-resolution neural activity for classifier training, and selectively processes informative channels via energy-efficient inference. Enhanced by multi-symptom biomarker extraction and a multi-class tree-structured neural network (NeuralTree) classifier, the SoC provides greater flexibility for a broader range of applications beyond seizure detection. Through efficient circuit implementation and circuit-algorithm co-design, this high-channel-count neural interface achieves a new class of energy-area efficiency.

This paper extends upon our prior work in \cite{shin2022256} and presents a review of the state-of-the-art, detailed description of the proposed circuits/algorithms, and more extensive benchtop and \emph{in-vivo} validation of the  SoC. The paper is organized as follows. Section~II provides a high-level description of the  256-channel SoC. Section III describes the system-level optimization of the AFE  and introduces a 256-channel time-division multiplexed (TDM) architecture with a two-step fast-settling mixed-signal DC servo loop (DSL). Sections IV and V detail the NeuralTree classification processor and the 16-channel high-voltage (HV) compliant neurostimulator, respectively. Benchtop and \emph{in-vivo} measurement results are demonstrated in Section VI. Finally, Section VII concludes the paper.

\begin{figure}[t]
  \centering
  \includegraphics[width=1\columnwidth]{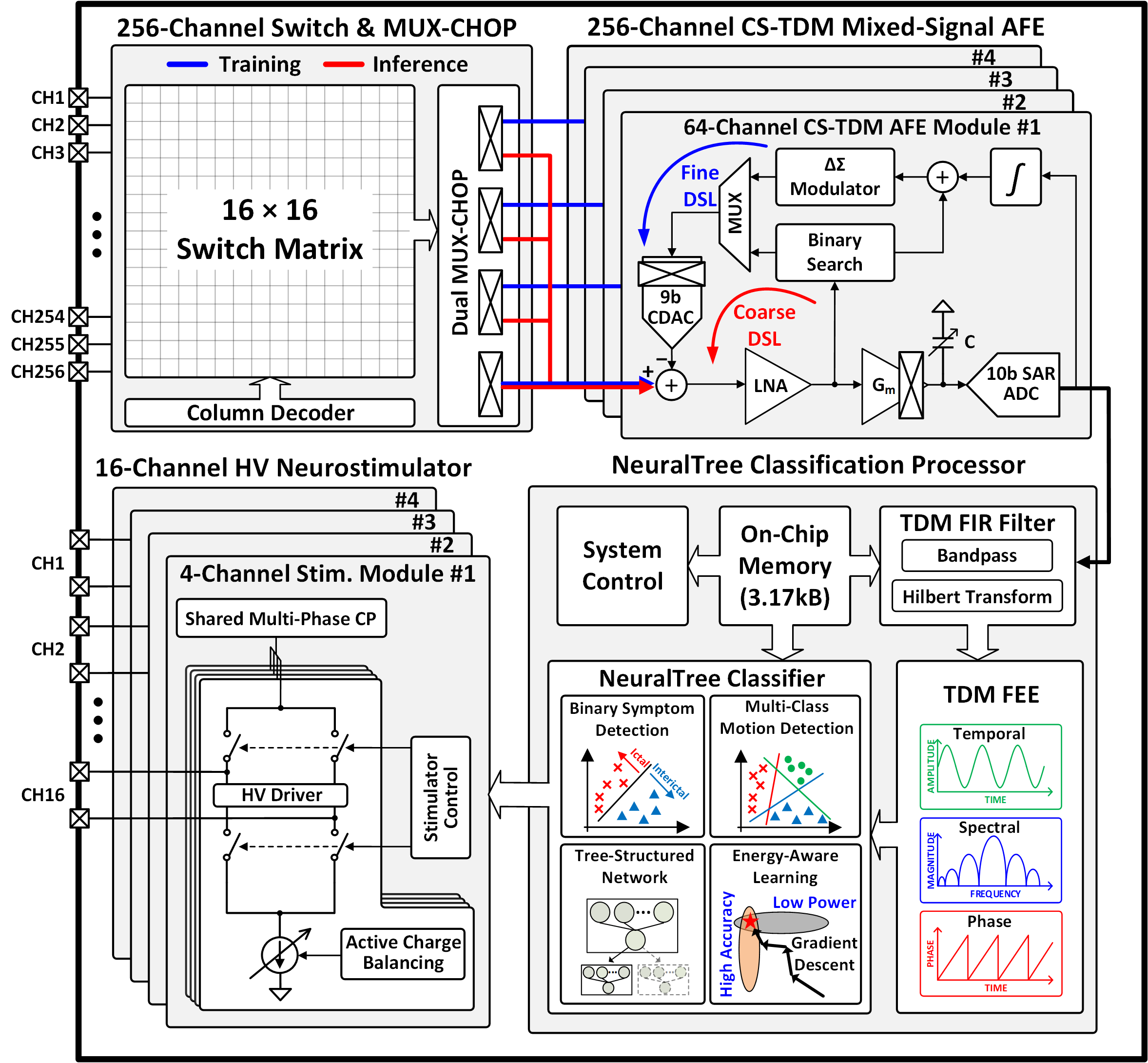}
  \vspace{-3mm}  
  \caption{Proposed 256-channel scalable, versatile closed-loop SoC architecture.}\vspace{-3mm}
  \label{f2_SoC}
\end{figure}

\vspace{-2mm}
\section{SoC Architecture}
Fig. \ref{f2_SoC} presents the architecture of the proposed 256-channel neural activity classification and closed-loop neuromodulation SoC. Four 64-channel chopper-stabilized time-division multiplexed (CS-TDM) AFE modules perform high-density multi-site neural recording to train the classifier. A 16$\times$16 switch matrix and row-multiplexing chopper (MUX-CHOP) connect the 256-channel neural inputs to the AFE modules for signal conditioning. In inference mode, the MUX-CHOP is configured such that any subset of 64 input channels can be selected and processed by the main AFE module. The input selection can change dynamically on a window-by-window basis according to the trained input sequence, to perform channel-selective inference. Following signal conditioning by the main AFE module, a finite impulse response (FIR) filter and feature extraction engine (FEE) compute neural biomarkers in temporal, spectral, and phase domains. Up to 64 multi-symptom biomarkers are extracted on demand in a patient- and disease-specific manner. The extracted biomarkers are passed to the NeuralTree classifier for neural activity classification. The flexible NeuralTree model supports both binary and multi-class classification. Upon detection of pathological brain states, a 16-channel HV compliant neurostimulator delivers charge-balanced biphasic current pulses to the brain to close the therapeutic loop.

The \emph{all}-\emph{in}-\emph{one} integration of high-density recording and stimulation channels, multi-symptom neural biomarkers, and ML intelligence can easily grow the hardware complexity of the system. To overcome this challenge, the proposed SoC employs various circuit-algorithm innovations and system-level hardware optimization techniques, which will be discussed in the remainder of this paper.

\begin{figure}[t]
  \centering
  \includegraphics[width=0.8\columnwidth]{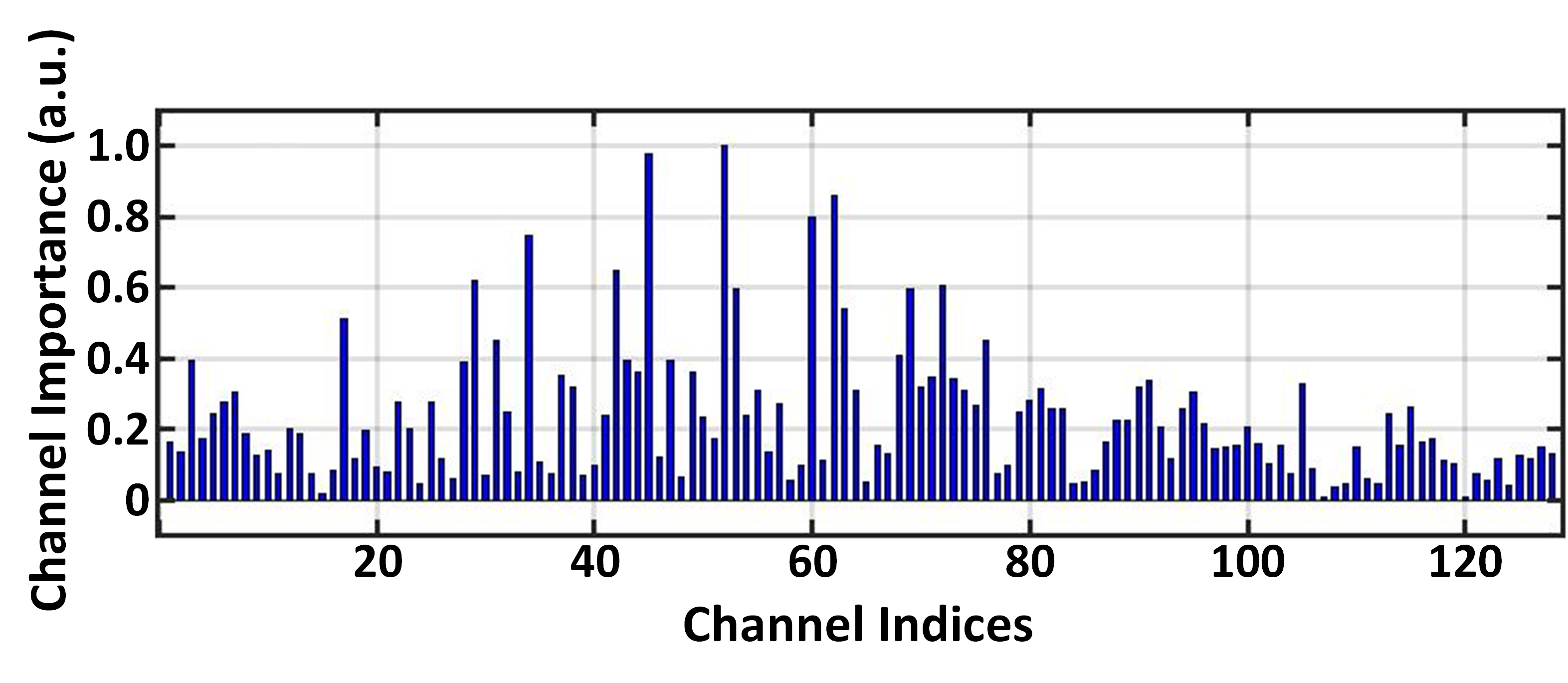}
  \vspace{-3mm}
  \caption{The relative importance of 128 iEEG channels recorded from an epilepsy patient and evaluated with 5-fold cross-validation.}\vspace{-3mm}
  \label{f3_importance}
\end{figure}

\vspace{-3mm}
\section{256-Channel Analog Front-End}\vspace{-1mm}
As the sensor count increases, the area constraint on the AFE  becomes more stringent and the complexity of the back-end signal processing also grows significantly. We tackle these challenges with an area-efficient TDM AFE with a channel-selective inference scheme. Noting that only subsets of input electrodes capture disease-relevant neural activity, the channel-selective approach can greatly reduce the hardware overhead during inference. To validate this concept, we trained a classifier on 128-channel intracranial electroencephalography (iEEG) recorded from an epilepsy patient \cite{ieeg} to assess the discriminative power of each channel. The NeuralTree classifier (detailed in Section. IV-C) was trained using two common types of seizure biomarkers (line-length and multi-band spectral energy) extracted from the 128 channels. The importance of each channel was then assessed based on the number of features  extracted during inference using 5-fold cross-validation. The non-uniform channel importance in Fig.~\ref{f3_importance} implies that high-density training followed by channel-selective inference can save the inference cost significantly while maintaining the classification accuracy.

\begin{figure}[t]
  \centering
  \includegraphics[width=1\columnwidth]{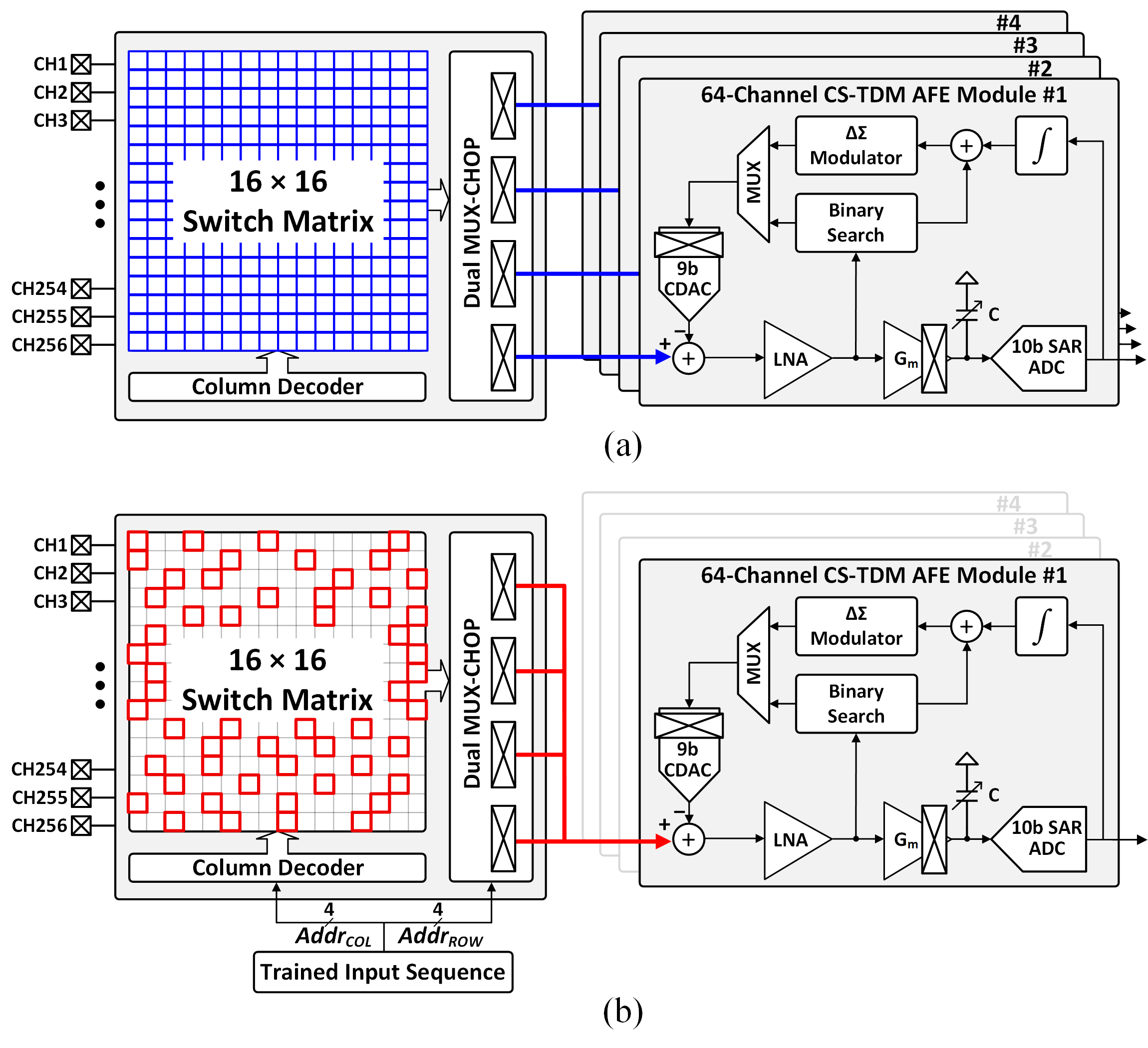}
  \vspace{-6mm}  
  \caption{The AFE configurations for (a) classifier training with high-density sensing, and (b) channel-selective inference.}\vspace{-3mm}
  \label{f4_AFE_config}
\end{figure}

Fig. \ref{f4_AFE_config} depicts the AFE configurations during classifier training and inference modes. In training mode, the four 64-channel CS-TDM AFE modules acquire 256-channel neural signals to exploit high-resolution brain activity information. The digitized 256-channel neural data are then used for offline classifier training, during which informative channel indices and feature types are identified. After loading the trained  parameters to an on-chip memory, the MUX-CHOP is configured to select any subset of up to 64 informative channels and connect them to the main AFE module via a shared input path. Here, the three auxiliary AFE modules are disabled to save system power. In the proposed NeuralTree model, the selected channels can change dynamically in each feature computation window to perform on-demand biomarker extraction, thus reducing the number of extracted features and enabling energy-efficient classification. However, the dynamic channel selection approach raises new concerns on electrode DC offset (EDO) cancellation that must be addressed.

\begin{figure}[t]
  \centering
  \includegraphics[width=0.8\columnwidth]{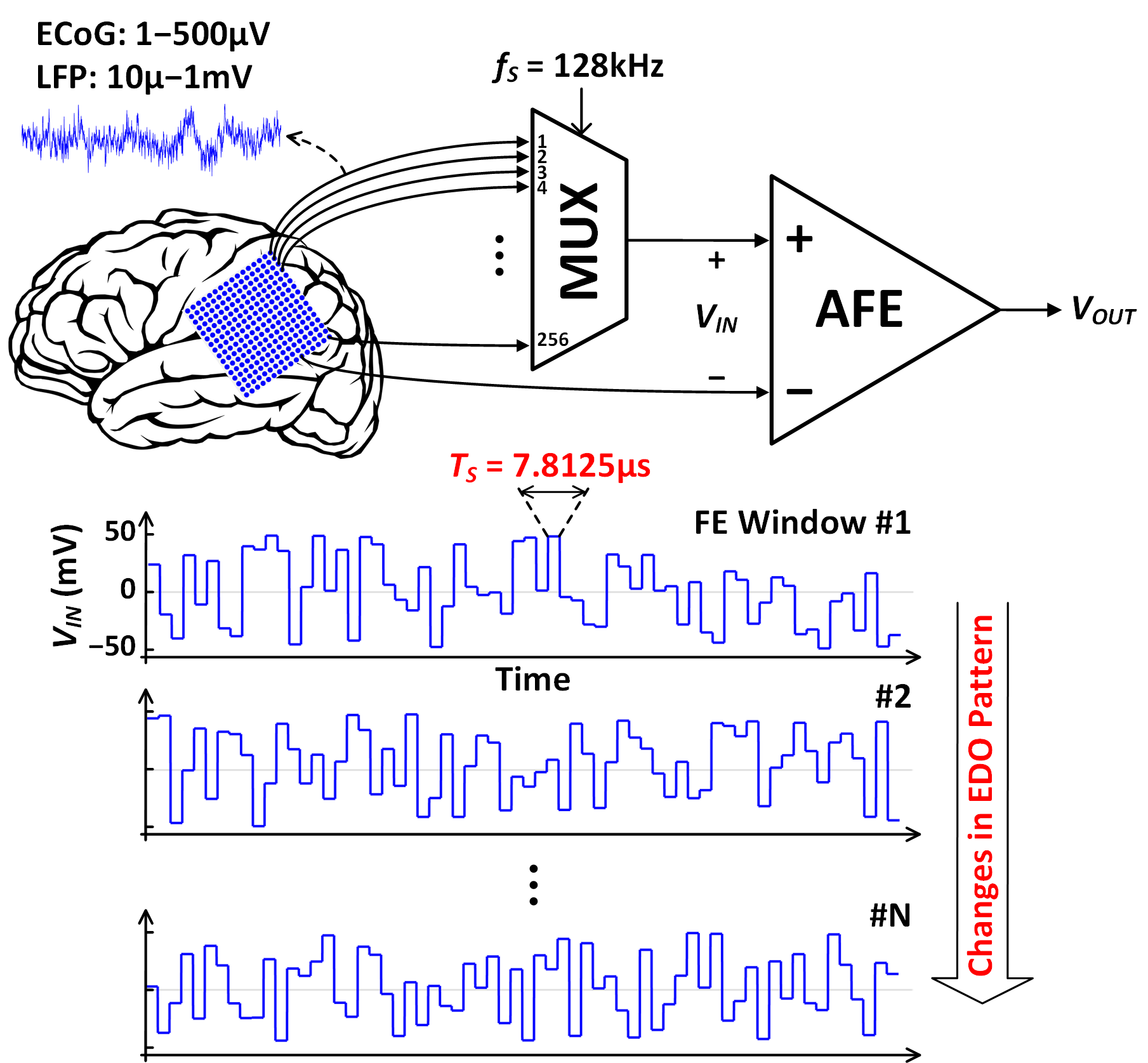}
  \vspace{0mm}
  \caption{Illustration of EDO fluctuations at the input of the proposed TDM AFE in channel-selective inference mode. Among 256 input channels, up to 64 channels with unique EDOs are multiplexed to the AFE in each feature extraction window. Therefore, the AFE must cancel the EDOs that change abruptly between successive channels and feature extraction windows.}\vspace{-3mm}
  \label{f5_EDO}
\end{figure}

\begin{figure*}[t]
  \centering
  \includegraphics[width=1.8\columnwidth]{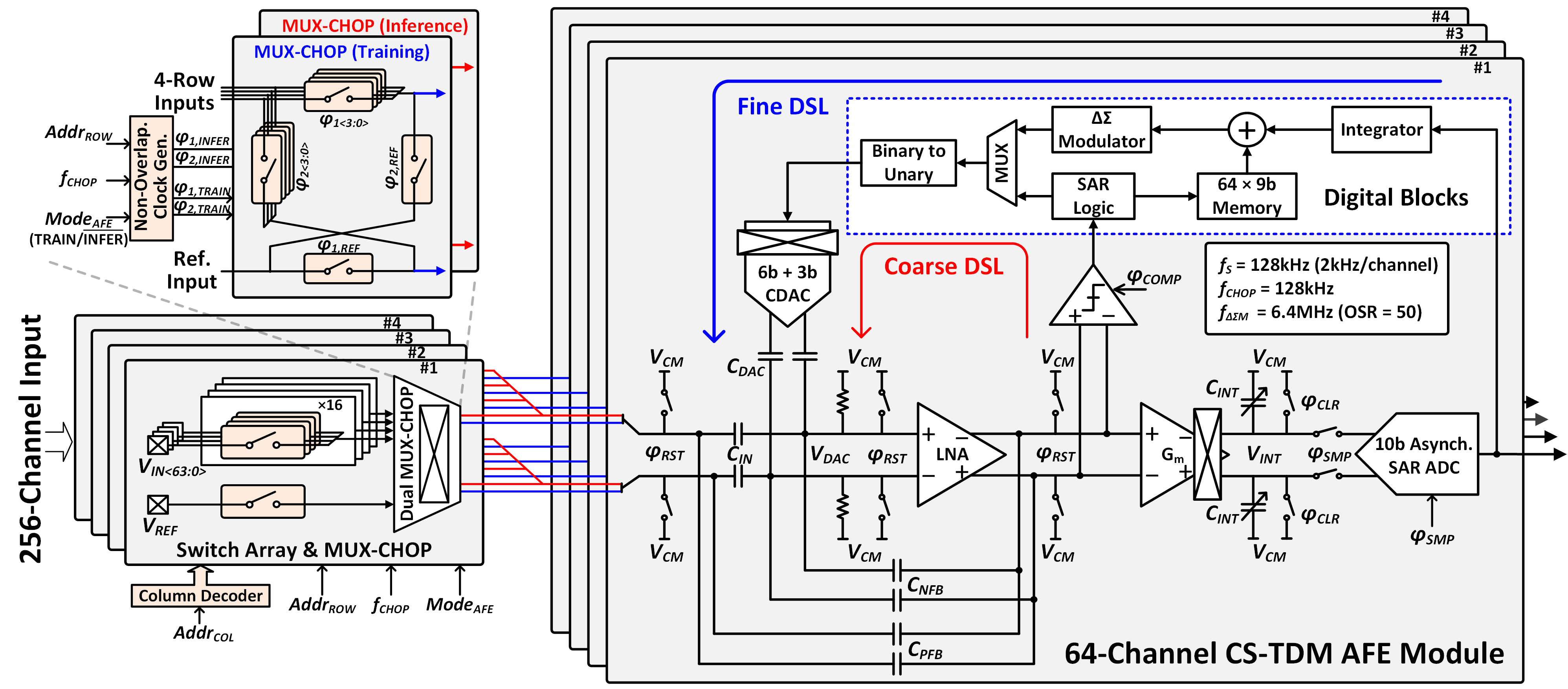}
  \vspace{-0mm}
  \caption{Modular architecture of the proposed 256-channel CS-TDM AFE with a two-step fast-settling mixed-signal DSL. In training mode, each AFE module sequentially conditions 64 input channels. In inference mode, the four MUX-CHOPs select up to 64 channels out of 256 and route them to the main AFE module via a shared input path drawn in red.}\vspace{-3mm}
  \label{f6_AFE}
\end{figure*}

\begin{figure}[t]
  \centering
  \includegraphics[width=0.6\columnwidth]{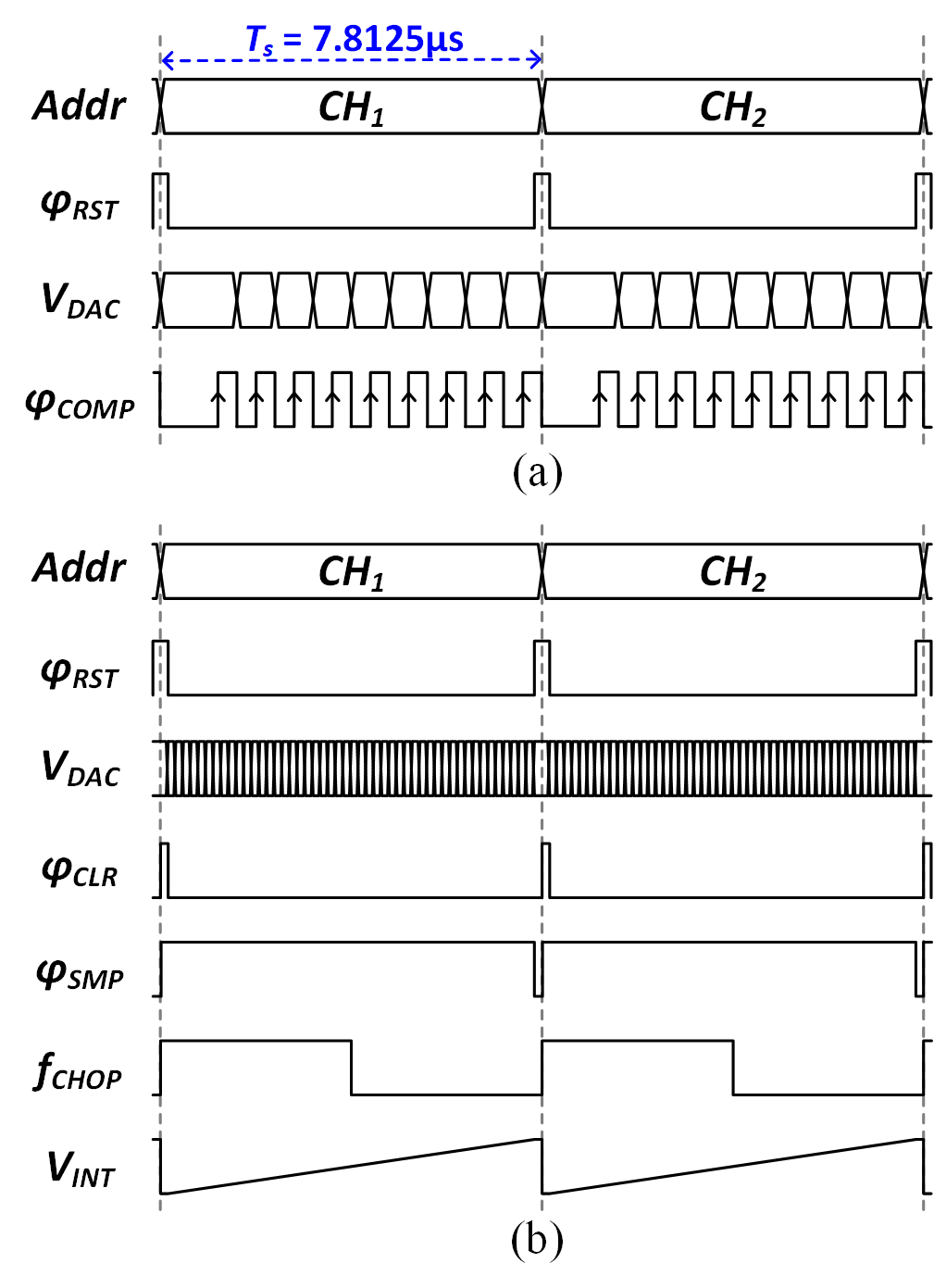}
  \vspace{-3mm}
  \caption{Timing diagram for the two-step (coarse/fine) DSL operation in channel-selective inference mode. (a) The coarse EDO cancellation using 9-bit binary search is performed in the first sampling period in each feature extraction window. (b) The $\Delta\Sigma$ fine loop for residual EDO cancellation operates for the rest of the feature extraction window.}\vspace{-3mm}
  \label{f7_DSL_timing}
\end{figure}

\vspace{-2mm}
\subsection{Challenges of Electrode DC Offset Cancellation}\vspace{-0.1mm}
Electrochemical polarization at the electrode-tissue interface develops DC offsets between electrodes \cite{cogan2008neural}, the magnitude of which can be as large as $\pm$50mV \cite{muller20110}. Each recording electrode in an array can develop a unique EDO with respect to a common reference electrode. When many electrodes with different EDOs are multiplexed to a single amplifier, these EDOs appear as a large signal that fluctuates at the multiplexing frequency, as shown in Fig. \ref{f5_EDO}. Digitizing small neural signals ($\sim$1$\upmu$\text{--}1mV) and significantly larger EDOs simultaneously would require a high-resolution ($\sim$16 bits) analog-to-digital converter (ADC), which is nontrivial to design in a compact and power-efficient way. A high-resolution ADC would also increase the hardware complexity of the subsequent filtering and  signal processing in the digital back-end (DBE). Another challenge imposed by the  channel-selective inference scheme is that the EDO pattern at the amplifier input changes between successive feature extraction windows (Fig. \ref{f5_EDO}). With a conventional large-time-constant DSL, the amplifier would saturate predominantly in each window ($\sim$1s), resulting in a significant loss of neural activity information.

The 16-channel SoC in \cite{altaf201516} multiplexed a low-noise amplifier (LNA) for every two channels to save chip area. Intermediate node voltages were stored on 1.5pF sampling capacitors for fast switching between channels with different EDOs. However, this analog S/H-based approach is area inefficient ($\sim$0.49mm$^2$/channel) and thus, it is not a viable option for a high-channel-count system. A mixed-signal coarse-fine DSL was reported in \cite{muller20110}. The coarse loop canceled large EDOs using binary search, while the fine loop suppressed residual offsets. Despite achieving a small area of 0.013mm$^2$/channel, the ADC-assisted binary search loop requires many samples to converge, making it inadequate for fast settling in the presence of abrupt EDO changes. Recently, hardware sharing via time-division multiplexing has been increasingly adopted to improve the area efficiency of high-channel-count AFEs. Specifically, to cancel EDOs between successive channels, several DSL designs have been reported, including binary search~\cite{sharma2019verification}, delta encoding \cite{uehlin20190}, and least-mean-square filtering \cite{fathy2021digitally}. While these AFEs can ultimately settle for a fixed input EDO pattern, none are compatible with channel-selective feature extraction scheme as the offset cancellation loops must re-settle in each window when a new set of inputs (with unknown EDO patterns) is fed to the AFE.    

\begin{figure*}[t]
  \centering
  \includegraphics[width=1.8\columnwidth]{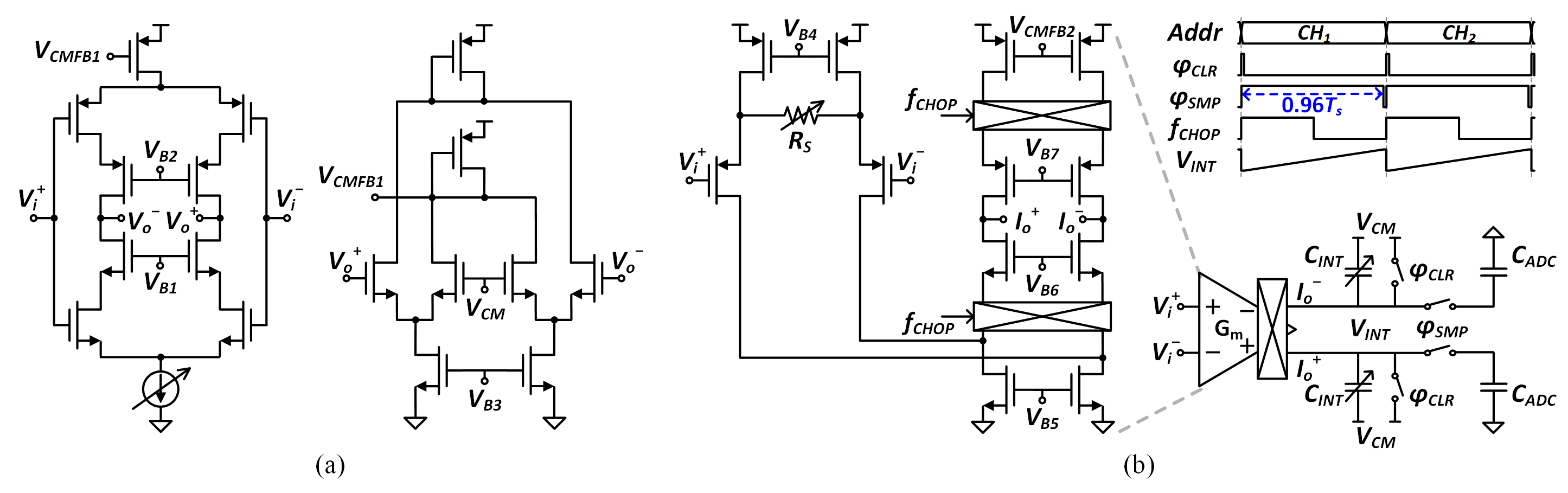}
  \vspace{-3mm}  
  \caption{Circuit implementations of the feedforward amplifiers: (a) core amplifier in the LNA stage and its common-mode feedback (CMFB) circuit, and (b) $\textit{G}_{\textit{m}}$-$\textit{C}$ integrator and the timing diagram of operation.}\vspace{-3mm}
  \label{f8_amplifiers}
\end{figure*}

\vspace{-2mm}
\subsection{Two-Step Fast-Settling Mixed-Signal DC Servo Loop}
To enable an area-efficient AFE implementation and address the EDO cancellation challenge, a 256-channel CS-TDM AFE with a two-step fast-settling mixed-signal DSL is proposed, as shown in Fig. \ref{f6_AFE}. The timing diagram of the two-step EDO cancellation process during inference is illustrated in Fig. \ref{f7_DSL_timing}. At the beginning of each feature extraction window, coarse offset cancellation using binary search is performed for 64 selected channels. A dynamic comparator detects the polarity of the LNA output, and a successive approximation register (SAR) logic subsequently updates the input code for a 9-bit capacitive digital-to-analog converter (CDAC). This process is repeated 9 times until the LNA output converges, and the EDO ($\leq$$\pm$50mV) is digitized at 9-bit resolution. The 9-bit EDO code is then stored into a register. This 64-channel coarse EDO cancellation is performed only during the first sampling period (7.8$\upmu$s/channel) of each window to enable fast settling and minimize neural data loss. Next, a fine loop is enabled to record neural signals and cancel any residual EDOs following coarse offset cancellation ($<$$\pm$0.2mV). A low-pass filtering digital integrator extracts undesired low-frequency signal components including residual EDOs from the ADC output. The pre-stored 9-bit EDO is added to the 9 most significant bits of the 19-bit integrator output. This newly formed 19-bit code is then $\Delta\Sigma$-modulated and fed back to the amplifier input via the 9-bit segmented (6-bit unary+3-bit binary) CDAC. Here, an oversampling ratio (OSR) of 50 increases the effective number of bits of the 9-bit CDAC to $\sim$17 bits to suppress the DAC quantization noise $<$1$\upmu$V \cite{muller2014minimally}. The output of the digital integrator can be bit-shifted to control the feedback gain for loop stability and adjust the highpass pole location in the closed-loop frequency response.

\vspace{-2mm}
\subsection{Low-Noise Amplifier and Anti-Aliasing Integrator}
The feedforward path of the AFE consists of a chopper-stabilized LNA and $\textit{G}_{\textit{m}}$-$\textit{C}$ integrator. The LNA is capacitively coupled to provide a precise, moderate closed-loop gain of 26dB ($\textit{C}_{\textit{IN}}$/$\textit{C}_{\textit{NFB}}$). The positive feedback loop ($\textit{C}_{\textit{PFB}}$) partially compensates for input impedance degradation due to chopping. The core amplifier in the LNA stage adopts an inverter-based current-reuse topology for improved noise efficiency \cite{song2013430nw} as shown in Fig. \ref{f8_amplifiers}(a). The complementary input pairs are biased in weak inversion ($\textit{g}_{\textit{m}}$/$\textit{I}_{\textit{D}}$ $\approx$ 25) and constructed using thick-oxide transistors to prevent gate leakage. The cascode transistors boost the open-loop gain to enable a precise closed-loop gain. They further mitigate the Miller effect of the input-pair gate-drain capacitance to prevent signal attenuation.

With a {$\Delta$}{$\Sigma$} frequency of 6.4MHz, the LNA bandwidth is set to 7MHz, which is much higher than the ADC sampling rate of 128kS/s. This high LNA bandwidth necessitates an anti-aliasing filter prior to digitization to avoid noise folding. Filtering is achieved by a charge-sampling $\textit{G}_{\textit{m}}$-$\textit{C}$ integrator, providing a \emph{sinc}-shaped frequency response with notches at integer multiples of the sampling frequency \cite{xu2000comparison, mirzaei2008analysis}. Fig.~\ref{f8_amplifiers}(b) presents the circuit implementation of the $\textit{G}_{\textit{m}}$-$\textit{C}$ integrator and the timing diagram of the charge-sampling operation. A folded-cascode amplifier  with source degeneration is implemented for improved linearity. The integration time extends to the 96\% of the sampling period, during which high-frequency {$\Delta$}{$\Sigma$} noise and chopper ripples are attenuated. The charge sampling approach relaxes the settling requirement of the amplifier \cite{xu2000comparison}, obviating the need for a power-consuming ADC buffer. The 3-bit resistor ($\textit{R}_{\textit{S}}$) and 5-bit capacitor ($\textit{C}_{\textit{INT}}$) banks provide an additional programmable gain (14\text{--}34dB) in the feedforward path. 

In a TDM front-end, samples of the current channel can be corrupted by previous channel residues, which manifests as inter-channel crosstalk. To prevent this, our CS-TDM AFE periodically resets the intermediate nodes along the feedforward path between successive channels. The \textit{kT}/\textit{C} noise resulting from the reset operation is up-modulated by the chopper and subsequently filtered out by the anti-aliasing $\textit{G}_{\textit{m}}$-$\textit{C}$ integrator.
\vspace{-2mm}
\subsection{Asynchronous Analog-to-Digital Converter}
A 10-bit SAR ADC digitizes the $\textit{G}_{\textit{m}}$-$\textit{C}$ integrator output at 128kS/s (64 channels). 
Following charge sampling in the $\textit{G}_{\textit{m}}$-$\textit{C}$ integrator, only a short time (312.5ns) is allocated to digitization. To avoid the use of an excessively high clock, we adopt an asynchronous SAR control with a single, master sampling clock \cite{liu201010}. A  9-bit binary-weighted charge-redistribution DAC with top-plate sampling and  monotonic-switching is implemented, using 2.3fF  metal-oxide-metal unit capacitors and bootstrapped switches \cite{liu201010}. An attenuation capacitor equal to the total 9-bit DAC capacitance is added to halve the effective input range of the ADC without an additional reference voltage generator \cite{harpe201126}. This approach relaxes the linearity and gain requirements of the preceding amplifiers with only a marginal area overhead, which is amortized across 64 channels in the proposed TDM AFE.
\vspace{-2mm}
\section{NeuralTree Classification Processor}
The high-level architecture of the proposed NeuralTree classification processor is presented in Fig. \ref{f2_SoC}. The configurable TDM FIR filter performs selective bandpass filtering (BPF) and Hilbert transformation (HT)  depending on the type of feature being extracted. Following signal filtering, the multi-symptom TDM FEE extracts up to 64 patient- and disease-specific neural biomarkers in temporal, spectral, and/or phase domains. The NeuralTree classifier uses the extracted feature vectors to perform top-down brain-state inference along the most probable path of the tree. The end-to-end TDM implementation enables a seamless integration of key building blocks without the need for demultiplexing, thus achieving a new class of scalability and energy efficiency. Hardware-friendly feature approximations and energy-aware  training algorithm further improve the NeuralTree's hardware efficiency, as detailed in this section. 

\begin{figure}[t]
  \centering
  \includegraphics[width=1\columnwidth]{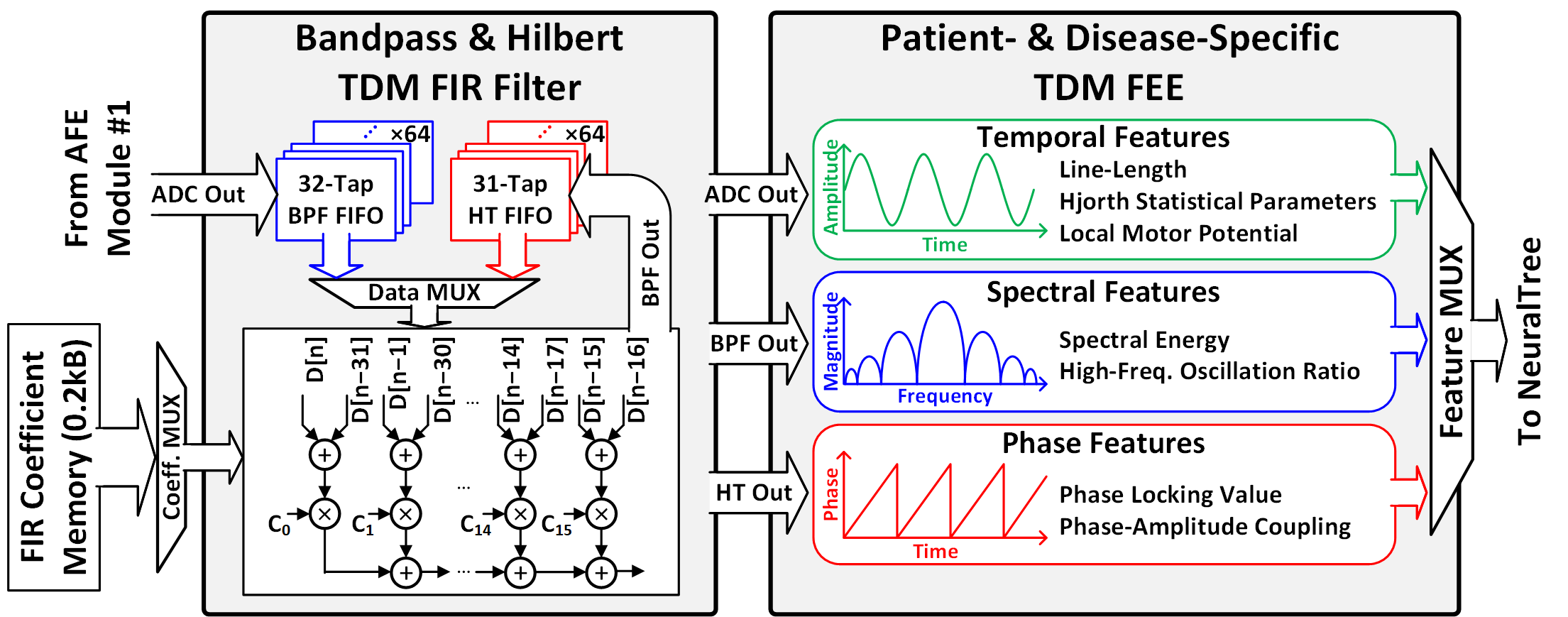}
  \vspace{-3mm}  
  \caption{Configurable TDM FIR filter and multi-symptom TDM FEE.}\vspace{0mm}
  \label{f9_FIR_FEE}
\end{figure}

\begin{figure}[t]
  \centering
  \includegraphics[width=1\columnwidth]{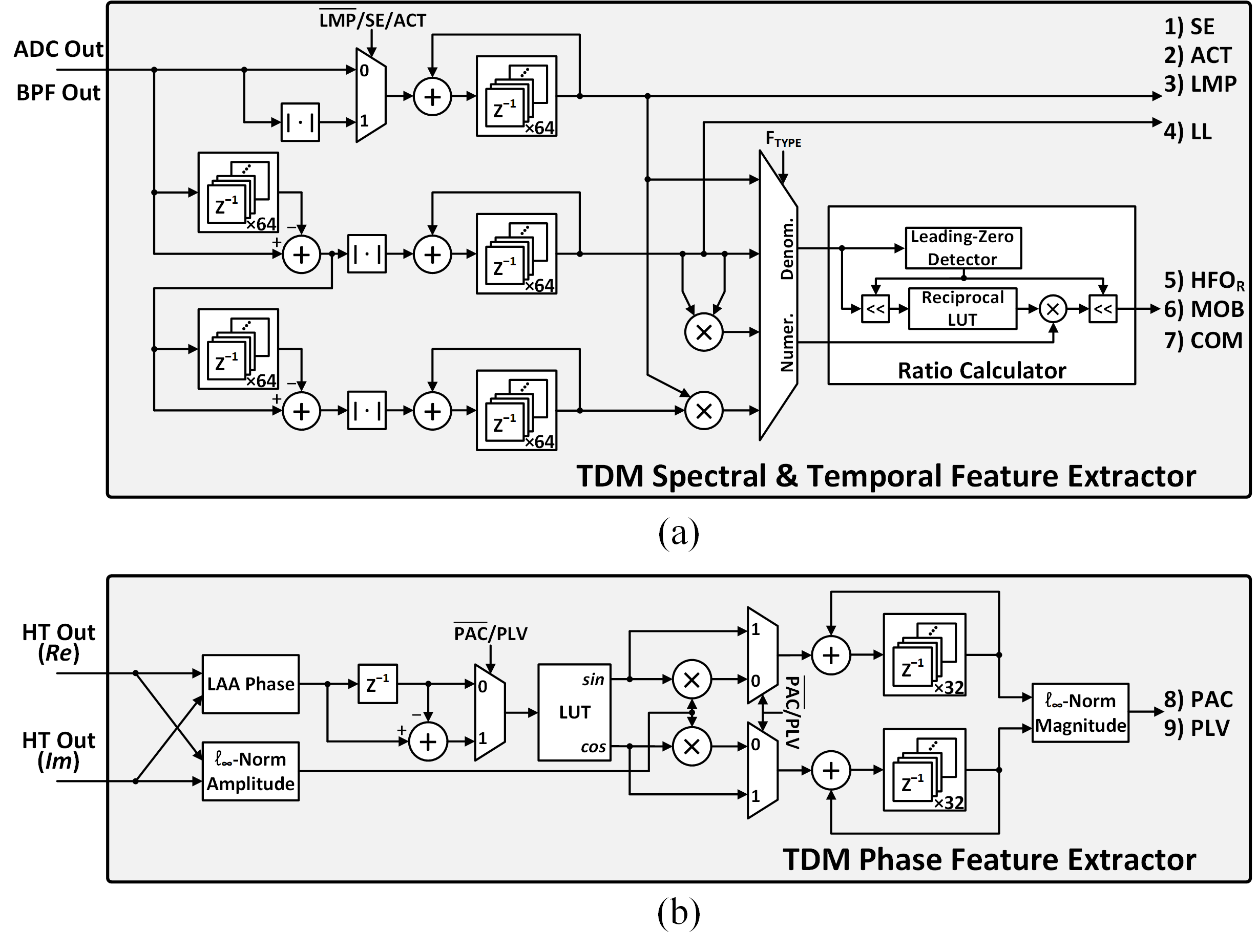}
  \vspace{-6mm}  
  \caption{Hardware implementations of the TDM FEE: (a) temporal and spectral feature extractor, and (b) phase feature extractor.}\vspace{-3mm}
  \label{f10_FEE}
\end{figure}

\vspace{-2mm}
\subsection{Bandpass Filtering and Hilbert Transform}
Fig. \ref{f9_FIR_FEE} depicts the block diagram of the TDM FIR filter that processes digitized neural signals. To save silicon area, a single set of arithmetic units is shared between 64 channels for 32-tap bandpass filtering \cite{altaf201516}. Band-specific FIR coefficients are retrieved from the on-chip memory and multiplexed into the multiplier array. The FIR filter can be reconfigured as a 31-tap Hilbert transformer to obtain analytic signals for instantaneous phase and amplitude extraction. The 64-channel BPF and HT register banks are clocked at 128kHz and selectively clock-gated depending on the feature type. For temporal feature extraction, the ADC output is directly fed to the FEE and the FIR is bypassed. 

\begin{table}[t]
  \centering  
  \vspace{-3mm}   
  \caption{Task-Specific Neural Biomarkers Integrated on the SoC}
  \includegraphics[width=0.9\columnwidth]{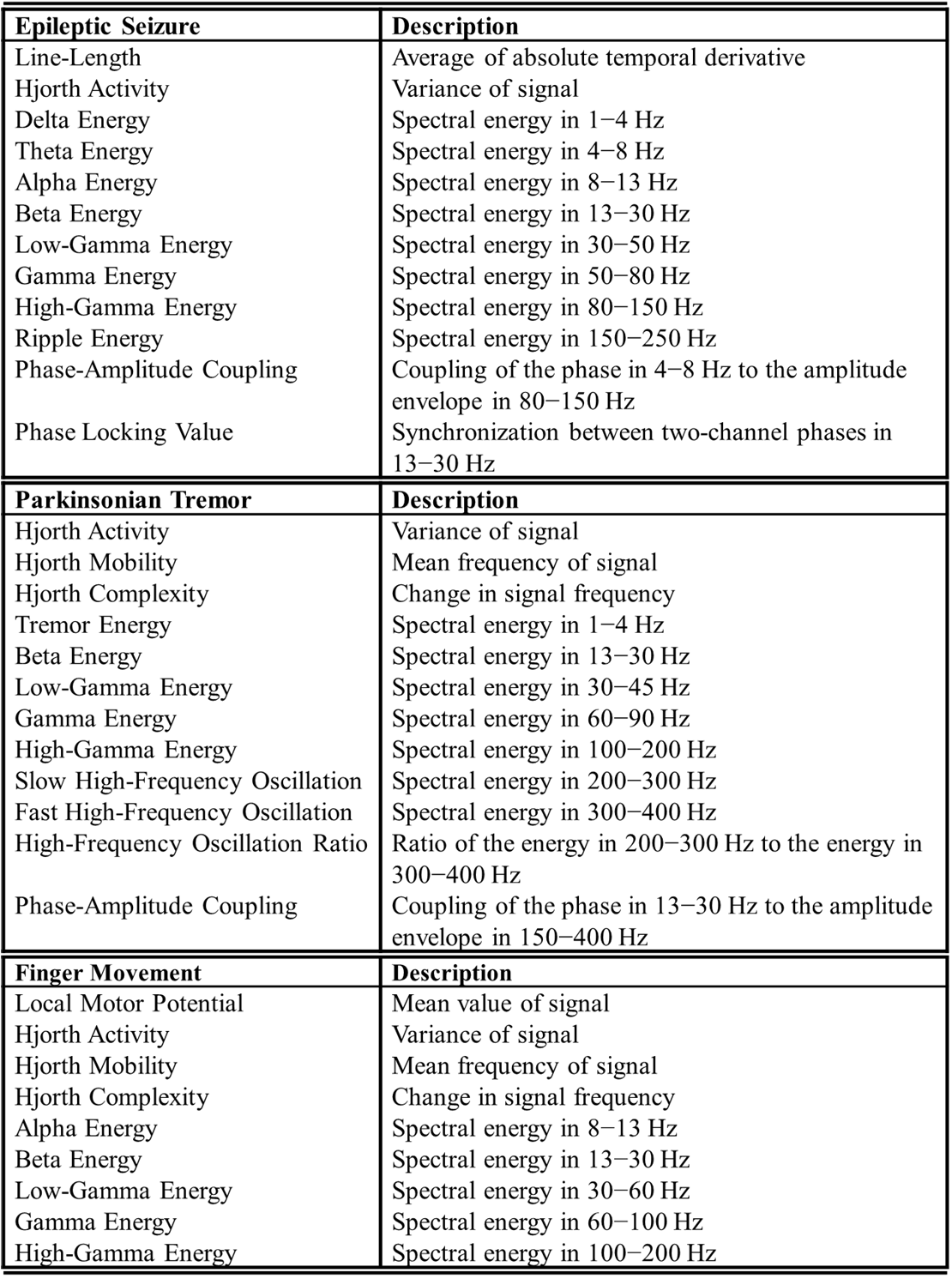}\vspace{-3mm}  
  \label{t1_feature_set}
\end{table}

\vspace{-2mm}
\subsection{Multi-Symptom Feature Extraction}
To enhance the versatility of the SoC for a broad range of neural classification tasks, the FEE integrates  multi-symptom neural biomarkers, as summarized in Table.~\ref{t1_feature_set}. Without careful design considerations, integrating such a broad range of biomarkers can be hardware intensive. The following subsections describe hardware-friendly feature approximation algorithms and circuit techniques that enable low-complexity, yet accurate feature extraction in the proposed SoC.

\subsubsection{Temporal features}
\indent Line-length (LL) increases in the presence of high-amplitude or high-frequency neural oscillations and has been among powerful biomarkers of epileptic seizures \cite{esteller2001line}. LL is defined as in (\ref{eq_ll}): 
\begin{equation}
\text{LL} = \frac{1}{N}\sum_{t=1}^{N} |{x_{t}-x_{t-1}}|\label{eq_ll}
\end{equation}
where N is the number of samples in a feature extraction window. Fig. \ref{f10_FEE}(a) presents the hardware implementation of the proposed TDM temporal feature extractor. For LL extraction, the absolute differences between successive samples are accumulated with two adders and one absolute value calculator. 

The Hjorth statistical parameters are highly correlated with tremor in Parkinson's disease \cite{yao2020improved} and used in BMIs for finger movement \cite{zhu2020resot} and gait \cite{shafiul2020prediction} decoding. The Hjorth activity (ACT), mobility (MOB), and complexity (COM) measure the variance, mean frequency, and frequency change of a signal, respectively, as defined below \cite{hjorth1970eeg}:
\begin{equation}
\text{ACT} = var(x) = \frac{1}{N}\sum_{t=1}^{N} ({x_{t}-\mu})^2 \label{eq_act}
\end{equation}
\begin{equation}
\text{MOB} = \sqrt{\frac{var(\Delta x)}{var(x)}} \label{eq_mob}
\end{equation}
\begin{equation}
\text{COM} = \sqrt{\frac{var(x) \cdot var({\Delta}^2 x)}{{var}^2 (\Delta x)}} \label{eq_com}
\end{equation}
where $\mu$, $\Delta x$, and ${\Delta}^2 x$ are the mean, first, and second derivatives of the signal $x$, respectively.

The three Hjorth parameters are difficult to efficiently compute  in their original form, due to the intensive multiplication and square root operations. Ref. \cite{goncharova1990changes} introduced a similar set of parameters, namely mean amplitude, mean frequency, and spectral purity index (SPI), in which the root of square operator is replaced by simple absolute value approximation. These new parameters are less intensive to compute while preserving a close relation to the measures of EEG amplitude and frequency. We adopt this approach to approximate the Hjorth features as in (\ref{eq_act2})\text{--}(\ref{eq_com2}), with a modification to the SPI parameter by taking its reciprocal, since it is better correlated with the original Hjorth complexity parameter:
\begin{equation}
\text{ACT} \approx \frac{1}{N}\sum_{t=1}^{N} |{x_{t}}| \label{eq_act2}
\end{equation}
\begin{equation}
\text{MOB} \approx \frac{\sum_{t=1}^{N} |{\Delta x_{t}}|}{\sum_{t=1}^{N} |{x_{t}}|} \label{eq_mob2}
\end{equation}
\begin{equation}
\text{COM} \approx \frac{(\sum_{t=1}^{N} |{x_{t}}|) \cdot (\sum_{t=1}^{N} |{{\Delta}^2 x_{t}}|)}{(\sum_{t=1}^{N} |{\Delta x_{t}}|)^2} \label{eq_com2}
\end{equation}

To calculate the approximated Hjorth features, the absolute values of the input and its first and second derivatives are accumulated selectively, as shown in Fig. \ref{f10_FEE}(a). For MOB and COM extraction, the subsequent multipliers and ratio calculator further process the accumulated derivatives to compute  features in  fractional form. The ratio calculator employs a reciprocal-multiply approach with bit shifting instead of a complex divider, as depicted in Fig.~\ref{f10_FEE}(a).   

Local motor potential (LMP) has been used as a low-complexity yet effective marker for motor intention decoding in BMIs \cite{stavisky2015high}. The LMP feature quantifies the mean value of a signal as defined in (\ref{eq_lmp}):
\begin{equation}
\text{LMP} = \frac{1}{N}\sum_{t=1}^{N} x_{t} \label{eq_lmp}
\end{equation}
The accumulation function can be performed by reusing the ACT extractor and bypassing the absolute value calculator, as shown in Fig. \ref{f10_FEE}(a).

\subsubsection{Spectral features}
\indent Spectral energy (SE) in multiple frequency bands of neural oscillations has been a commonly used biomarker in epilepsy \cite{altaf201516, o2018recursive, huang20191, chua20211, zhang2022patient, shoaran2018energy, zhu2020resot}, Parkinson's disease \cite{yao2020improved, yao2018resting}, and BMIs \cite{so2014subject, yao2022fast}. As a measure of signal power integrated over time, the SE can be defined in the discrete-time domain, as follows: 
\begin{equation}
\text{SE} = \frac{1}{N}\sum_{t=1}^{N} {x_{\text{BAND},t}}^2 \label{eq_se}
\end{equation}
where $x_{\text{BAND},t}$ indicates the bandpass-filtered neural signal. 

A common approximation method to avoid the square operation is to take the absolute output of the bandpass filter. The 16-channel EEG processor in \cite{altaf201516} demultiplexed the output of the TDM FIR filter to 112 signal paths (16 channels$\times$7 bands) to calculate 112 SE features in parallel. This approach requires an equal number of multi-bit adders and absolute value calculators with significant area overhead. To save chip area, the TDM spectral feature extractor in Fig. \ref{f10_FEE}(a) directly receives the BPF output as the input without demultiplexing, and extracts up to 64 SE features using a single adder. The area efficiency is further improved by reusing the hardware already implemented for ACT and LMP extraction.

High-frequency ($>$200Hz) oscillations (HFOs) are prominent features in Parkinson's disease \cite{ozkurt2011high} and epilepsy ($>$80Hz) \cite{jacobs2012high}.  For instance, \cite{hirschmann2016parkinsonian} reported  the energy ratio between the slow  ($\text{HFO}_\text{1}$, 200\text{--}300Hz) and fast HFO ($\text{HFO}_\text{2}$, 300\text{--}400Hz) as an indicator of rest tremor in PD:
\begin{equation}
\text{HFO}_\text{R} = \frac{\sum_{t=1}^{N} {x_{\text{HFO}_\text{1},t}}^2}{\sum_{t=1}^{N} {x_{\text{HFO}_\text{2},t}}^2} \label{eq_hforatio}
\end{equation}
The SE extractor is reused to calculate the slow and fast HFOs, while the ratio between the two is computed using the ratio calculator shared with the Hjorth feature extractor.

\subsubsection{Phase features}
\indent Different brain regions communicate with each other through neuronal oscillations. Abnormal cross-regional  synchronization of neural oscillations can indicate disease-related pathological states in neurological and psychiatric disorders. In epilepsy, spatial and temporal changes in cross-channel phase synchronization, quantified by phase locking value (PLV), play as a key indicator of  seizure state \cite{mormann2000mean}. Phase-amplitude coupling (PAC) is another mechanism for within- and cross-regional brain communication. PAC quantifies the degree to which the low-frequency neural oscillatory phase modulates the amplitude of high-frequency oscillations~\cite{canolty2010functional}. Excessive PAC has been observed in disorders such as epilepsy \cite{guirgis2013role}, Parkinson's  \cite{de2013exaggerated}, and depression~\cite{olbrich2014functional}.

Measuring PLV and PAC  requires Hilbert transformation to obtain analytic signals followed by several complex computations such as extraction of instantaneous phase and amplitude, trigonometric functions, and magnitude computation, as shown in (\ref{eq_plv}) and (\ref{eq_pac}):
\begin{equation}
\text{PLV} = \frac{1}{N}\sqrt{(\sum_{t=1}^{N} sin \Delta \theta_{t})^2+(\sum_{t=1}^{N} cos \Delta \theta_{t})^2}  \label{eq_plv}
\end{equation}
\begin{equation}
\text{PAC} = \frac{1}{N}\sqrt{(\sum_{t=1}^{N} A_{t}sin \theta_{t})^2+(\sum_{t=1}^{N} A_{t}cos \theta_{t})^2}  \label{eq_pac}
\end{equation}
where $\Delta \theta_{t}$ in (\ref{eq_plv}) is the cross-channel phase difference, and $\theta_{t}$ and $A_{t}$   in (\ref{eq_pac}) are the modulating phase and modulated amplitude envelope, respectively.  

The SoCs in \cite{abdelhalim201364, o2018recursive} employed multiple COordinate Rotation DIgital Computer (CORDIC) processors to compute these non-linear functions, consuming an excessive amount of power ($>$200$\upmu$W). Alternatively, Fig. \ref{f10_FEE}(b) depicts the proposed TDM phase feature extractor \cite{shin2022256}. With band-specific analytic signals (\textit{Re} and \textit{Im}) as inputs, the instantaneous phase is approximated using a linear arctangent approximation (LAA) algorithm \cite{rajan2006efficient} followed by look-up table (LUT)-based error correction \cite{cicc}. The \emph{$l_\infty$}-norm is used to approximate the amplitude envelope of high-frequency oscillations in PAC, as well as the magnitude computations  in (1) and (2). 
The TDM phase feature extractor can compute up to 32 PLV/PAC features on demand in a compact area of 0.033mm$^2$, performing a higher degree of multiplexing compared to the architecture in \cite{cicc}.

\begin{figure}[t]
  \centering
  \includegraphics[width=0.7\columnwidth]{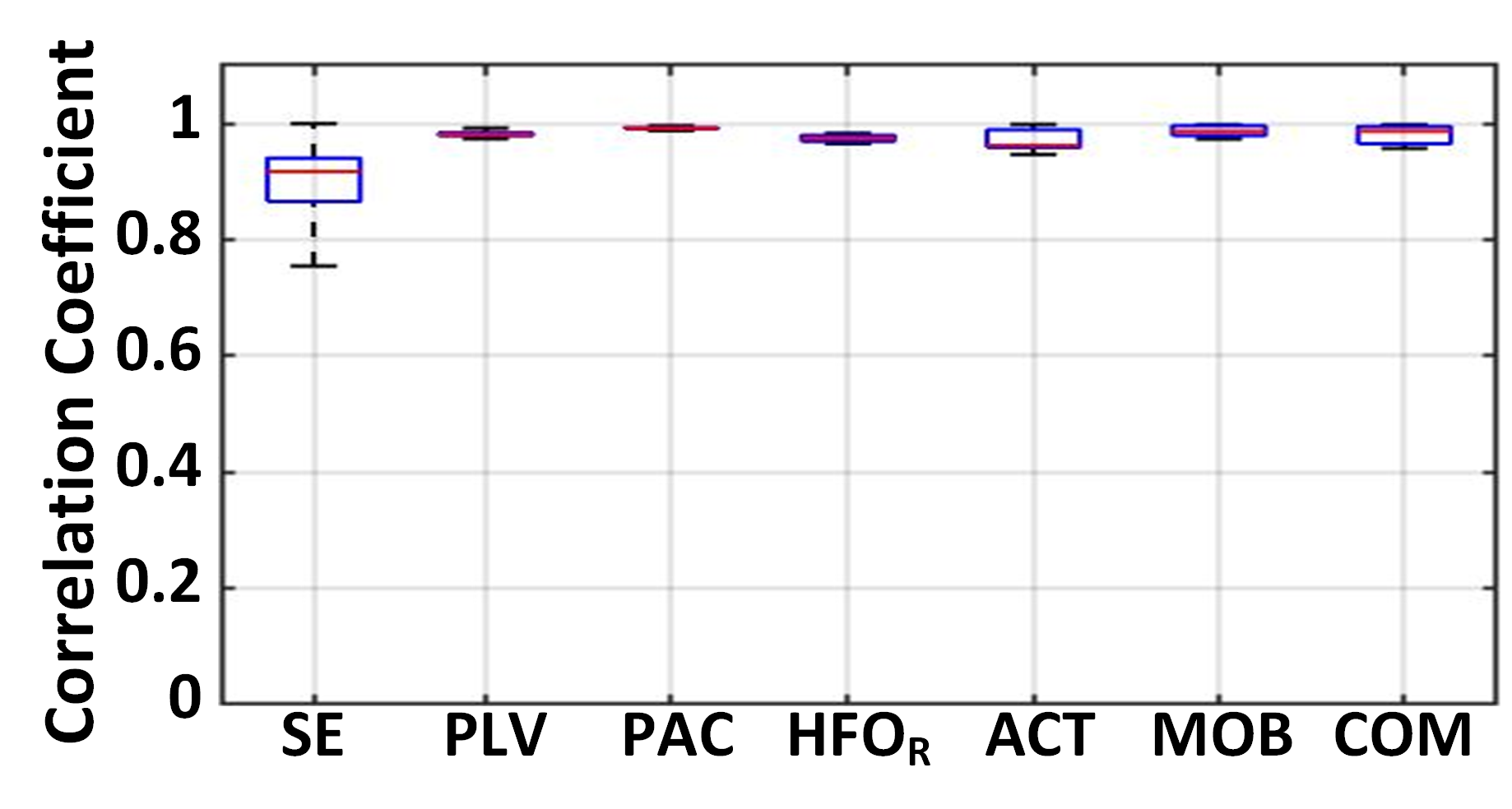}
  \vspace{-3mm}  
  \caption{Boxplot of the Pearson correlation coefficients between the ideal and approximated features.}\vspace{-3mm}
  \label{f11_correlation}
\end{figure}

To evaluate the accuracy of the proposed feature approximation algorithms, we analyzed the Pearson correlation coefficient between the ideal and approximated features in MATLAB. The phase and 8-band SE features were extracted from an epilepsy iEEG  dataset \cite{ieeg}, while a PD local field potential (LFP)  dataset \cite{yao2020improved} was used to compute the HFO ratio and Hjorth features. The boxplot of correlation coefficients in Fig.~\ref{f11_correlation} shows that the approximated features are highly correlated with their ideal counterparts, exhibiting median correlations above 0.9.

Thanks to feature approximations and hardware sharing, the proposed multi-symptom FEE  occupies a small silicon area of 0.12mm$^2$, even with the complex features integrated. Aggressive hardware sharing among different feature calculators is possible thanks to the on-demand TDM  scheme, where only selected features are consecutively extracted. With a 128kHz clock, the FEE can generate any combination of up to 64 neural biomarkers in a programmable feature extraction window (0.25\text{--}2s). Any unused hardware units are selectively clock- and data-gated to reduce dynamic power dissipation. 
\vspace{-2mm}
\subsection{Energy-Aware, Low-Complexity NeuralTree Model}
Deployment of ML algorithms in closed-loop neural interfaces can provide a more accurate and personalized treatment option compared to  conventional approaches with manual feature thresholding \cite{yoo2021neural}. However, stringent area and power constraints on implantable devices make it  challenging to integrate ML models with high computational  and memory requirements. These constraints become  more restrictive as the demand for higher channel counts continues to grow.

Decision tree (DT)-based ML models are becoming popular for their lightweight inference and low memory utilization. The 32-channel seizure detection classifier in \cite{shoaran2018energy} employed an ensemble of 8 eXtreme Gradient-Boosted (XGB) DTs. A 41.2nJ/class DBE energy efficiency  in 1mm$^2$ area was achieved, thanks to the on-demand feature extraction and sequential node processing. The 8-channel seizure detection SoC with 1024 AdaBoosted trees in \cite{o202026} reported low-complexity spectral feature extraction with bit-serial processing,  and achieved a 36nJ/class DBE energy efficiency. However, a  drawback of conventional DT-based classifiers is that the number of trees and signal processing units can significantly increase as the classification task becomes more complex. 

\begin{figure}[t]
  \centering
  \includegraphics[width=1\columnwidth]{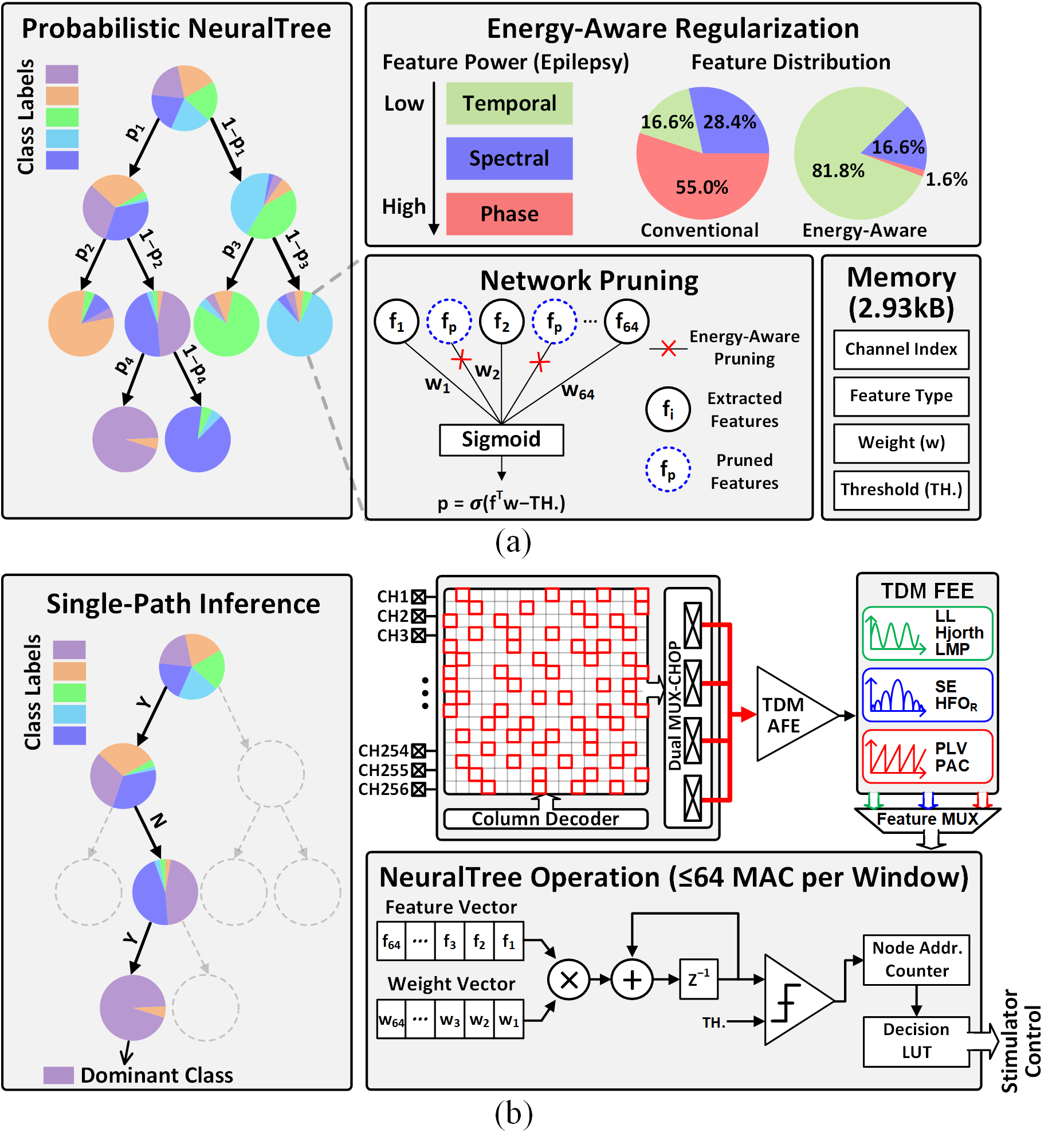}
  \vspace{-6mm}  
  \caption{NeuralTree classifier: (a) probabilistic NeuralTree trained with energy-aware regularization and network pruning, and (b) the NeuralTree hardware implementation and  system operation under the proposed single-path channel-selective inference scheme.}\vspace{-3mm}
  \label{f12_NeuralTree}
\end{figure}

\begin{figure*}[t]
  \centering
  \includegraphics[width=1.6\columnwidth]{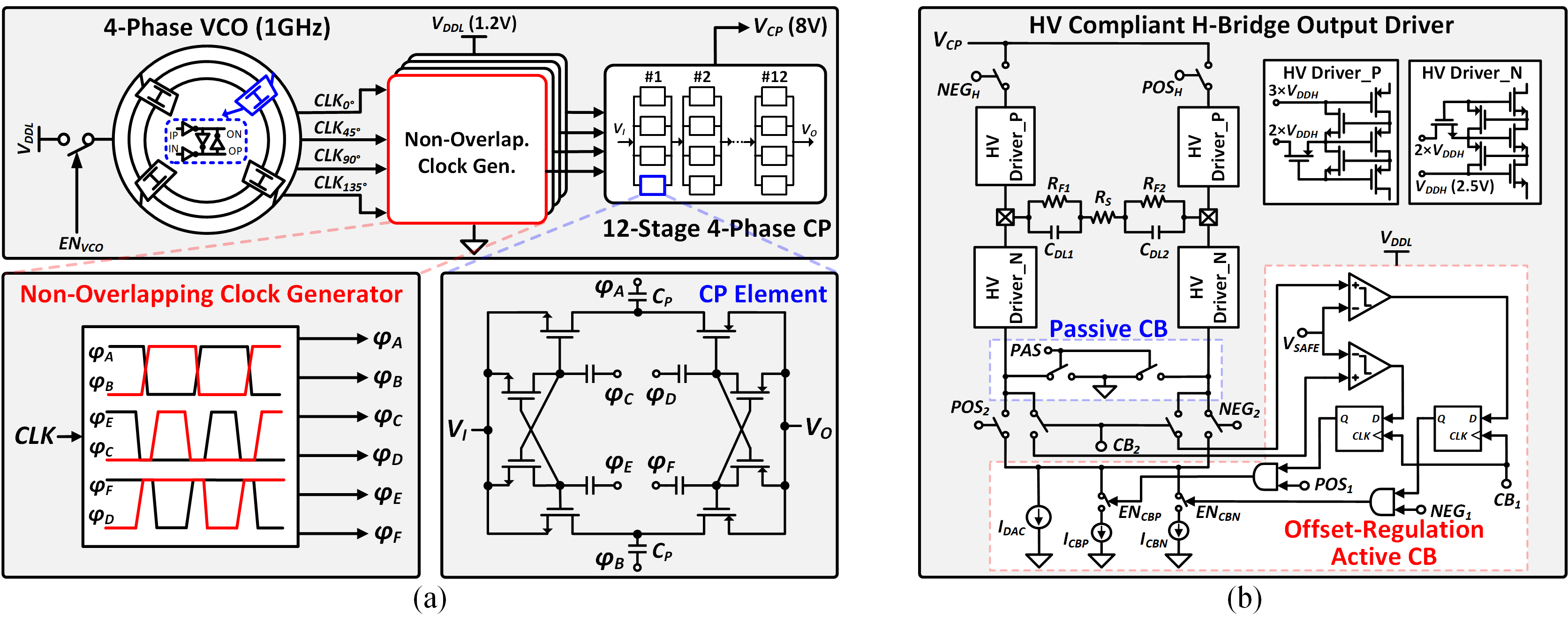}
  \vspace{-3mm}  
  \caption{Architecture of the HV compliant neurostimulator: (a) 12-stage 4-phase charge pump with high-frequency clocking, and (b) HV compliant H-bridge output driver with charge balancing.}\vspace{-3mm}
  \label{f13_STIM}
\end{figure*}

To enhance the hardware efficiency of the on-chip classifier, we propose a hierarchical NeuralTree model. Fig.~\ref{f12_NeuralTree}(a) illustrates a graphical representation of the NeuralTree classifier trained with a probabilistic routing scheme \cite{zhu2020resot}. Unlike conventional axis-aligned binary decision trees, the NeuralTree is a single oblique tree with probabilistic splits. With internal nodes represented by two-layer neural networks, the probability of splits in each internal node is computed using the sigmoid function. The feature vector $\boldsymbol{x_i}$ is routed to each leaf node $l$ with a probability $p(l|\boldsymbol{x_i};\boldsymbol{\theta})$, where $\boldsymbol{\theta}$ indicates the internal node parameters. The leaf nodes are parameterized by the class probability $\boldsymbol{\phi}$. The probability of $\boldsymbol{x_i}$ belonging to class $y_i$ can be expressed as:
\begin{equation}
p(y_i | \boldsymbol{x_i};\boldsymbol{\omega})=\sum_{l=1}^{L} p(l|\boldsymbol{x_i};\boldsymbol{\theta}) \boldsymbol{\phi_{l, y_i}} \label{eq_infer}
\end{equation}
\noindent where $\boldsymbol{\omega}=\boldsymbol{\theta} \cup \boldsymbol{\phi}$ indicates the trainable weights in the NeuralTree, and $L$ is the number of leaf nodes.
In the training process, we simply minimize the cross-entropy loss on the training data:
\begin{equation}
\min_{\boldsymbol{\omega}} \sum_{i=1}^{N} -\log p(y_i | \boldsymbol{x_i};\boldsymbol{\omega}). \label{eq_objective}
\end{equation}

The probabilistic NeuralTree is compatible with gradient-based optimization, which allows hardware-efficient model compression techniques such as weight pruning and fixed-point quantization. To enable neural activity inference with optimal energy-accuracy trade-off, the NeuralTree employs energy-aware regularization \cite{zhu2019cost}. Here, the power consumed for feature computation is added to the objective function as a regularization term in the training objective. Specifically, we define the energy-aware regularization term as: 
\begin{equation}
\Psi_{energy} =  \sum_{i=1}^{I} p_{i} \sum_{j=1}^{D} \beta_j |\theta_{i, j}| \label{eq_energy}
\end{equation}
\noindent where $I$ and $D$ represent the number of internal nodes and feature types, respectively, and $\beta$ is the normalized power cost for each feature type estimated using Synopsys PrimeTime. We use $p_i$ to represent the probability of visiting the internal node $i$ and $\theta_{i, j}$ for the weight associated with feature $j$ at node $i$. Combining (\ref{eq_objective}) and (\ref{eq_energy}), we derive an energy-aware objective function, which seeks to minimize the classification error as well as the energy consumption during  inference:
\begin{equation}
\min_{\boldsymbol{\omega}} \sum_{i=1}^{N} -\log p(y_i | \boldsymbol{x_i};\boldsymbol{\omega}) + C \cdot \Psi_{energy}(\boldsymbol{x_i};\boldsymbol{\omega}). \label{eq_train}
\end{equation}

We introduce a hyperparameter $C$ to control the energy-accuracy trade-off and penalize   power-demanding features. The proposed NeuralTree is trained using TensorFlow  \cite{abadi2016tensorflow}  with Adam  optimizer (learning rate: 0.001) \cite{kingma2014adam}. Validated on the iEEG epilepsy dataset \cite{ieeg}, the energy-efficient regularization saves 64$\%$ power in filtering and feature extraction during inference, with only a marginal accuracy loss  ($<$2$\%$). Following  regularization, network pruning is performed to compress the tree structure by reducing the number of extracted features (maximum 64 per node). 
Thanks to  network pruning and fixed-point weight quantization (12 bits), the trained parameters of the compressed NeuralTree  require only 2.93kB of memory. Through  energy-efficient regularization and network pruning, a set of features that lead to optimal energy-accuracy trade-off is extracted in a patient- and disease-specific manner. 

Considering that most samples are routed with high certainty during training, the inference can be performed through top-down conditional computations, as depicted in Fig. \ref{f12_NeuralTree}(b). By reusing a single multiply-and-accumulate (MAC) unit and a comparator, the standalone NeuralTree occupies a significantly smaller area compared to large tree ensembles. In addition to conventional binary classification tasks (e.g., seizure or tremor detection), the NeuralTree further supports multi-class classification tasks such as finger movement detection. This additional functionality is achieved with only a marginal memory overhead  to store multi-bit class labels in the decision LUT. As a proof of concept, in a 6-class finger movement classification task on a human ECoG dataset \cite{miller2012human}, the simulated NeuralTree decoded finger movements with 73.3\% accuracy. 

During inference, the NeuralTree performs sequential node processing  along the most probable path, thus reducing the  number of computed features  and weighted summations. The NeuralTree is clocked at 128kHz but only activated during the last 64 clock cycles in each feature extraction window to perform 64 feature-weight MAC operations. The lightweight channel-selective inference  coupled with energy-aware learning considerably enhances the model's energy efficiency and  scalability.
\vspace{-2mm}
\section{High-Voltage Compliant Neurostimulator}
A single chip integration of neurostimulator with  other building blocks is desired to reduce the size of the implantable device and interconnection complexity. However, depending on the electrode impedance and stimulation  amplitude, the voltage compliance required at the electrode-tissue interface can exceed the gate-oxide breakdown limits in standard CMOS processes \cite{uehlin2020single}. To facilitate a seamless integration of the closed-loop neuromodulation system in a standard low-power CMOS process, a 16-channel neurostimulator is implemented with a stacked high-voltage compliant architecture. 

Fig. \ref{f13_STIM}(a) presents the 12-stage 4-phase charge pump (CP) that generates an 8V supply for the current drivers. Each stage in the 4-phase CP consists of four elements in parallel, thus reducing output ripples by a factor of four without the need for large output capacitors \cite{van200982}. To minimize the reversion loss in the CP, flying capacitors ($\textit{C}_{\textit{P}}$) are controlled by a non-overlapping clock generator. The ring VCO generates a high-frequency clock at 1GHz, which allows the use of small flying capacitors (0.8pF) for improved area efficiency \cite{uehlin2020single}. To further save chip area, a single CP is shared among four stimulation channels that are individually addressable.

Fig. \ref{f13_STIM}(b) shows the architecture of the stacked H-bridge output driver. A single current sink ($\textit{I}_{\textit{DAC}}$) is used in both anodic and cathodic phases to reduce charge mismatch. To ensure precise charge balancing (CB), an offset-regulation active CB technique combined with passive discharging is employed~\cite{noorsal2011neural}. The residual voltage at the electrode is monitored following biphasic stimulation. If the residual voltage exceeds the safety level ($\pm\textit{V}_{\textit{SAFE}}$), the active CB is enabled to provide an additional current such that the two stimulation phases are balanced. Following  active CB, any residual charges on the two electrodes are discharged to the ground. To provide sufficient flexibility for various applications, the stimulation parameters such as  current amplitude, pulse width, and frequency are programmable.

\begin{figure}[t]
  \centering
  \includegraphics[width=0.8\columnwidth]{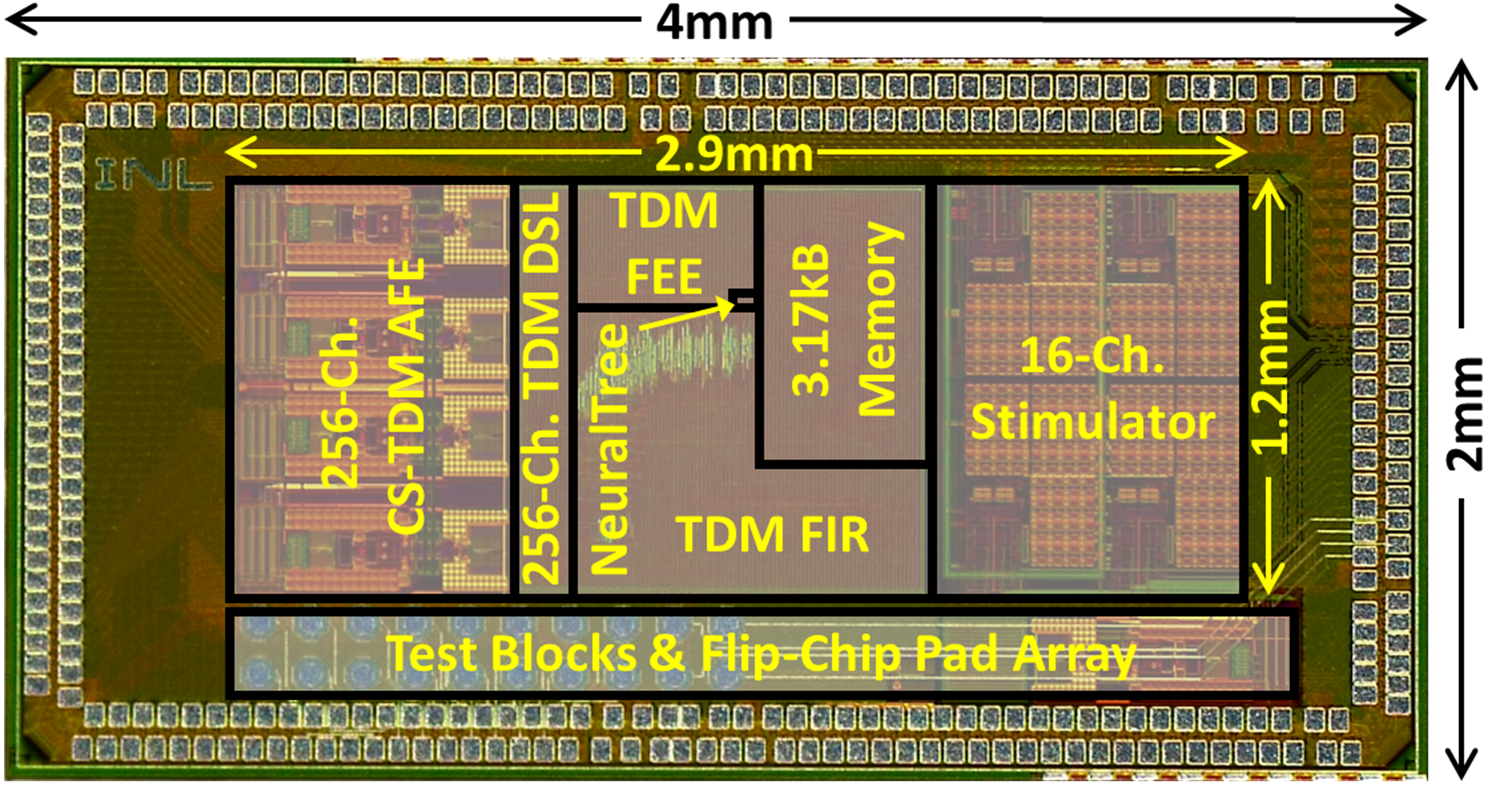}
  \vspace{0mm}  
  \caption{Chip micrograph.}\vspace{-3mm}
  \label{f14_chip}
\end{figure}

\begin{figure}[t]
  \centering
  \includegraphics[width=1\columnwidth]{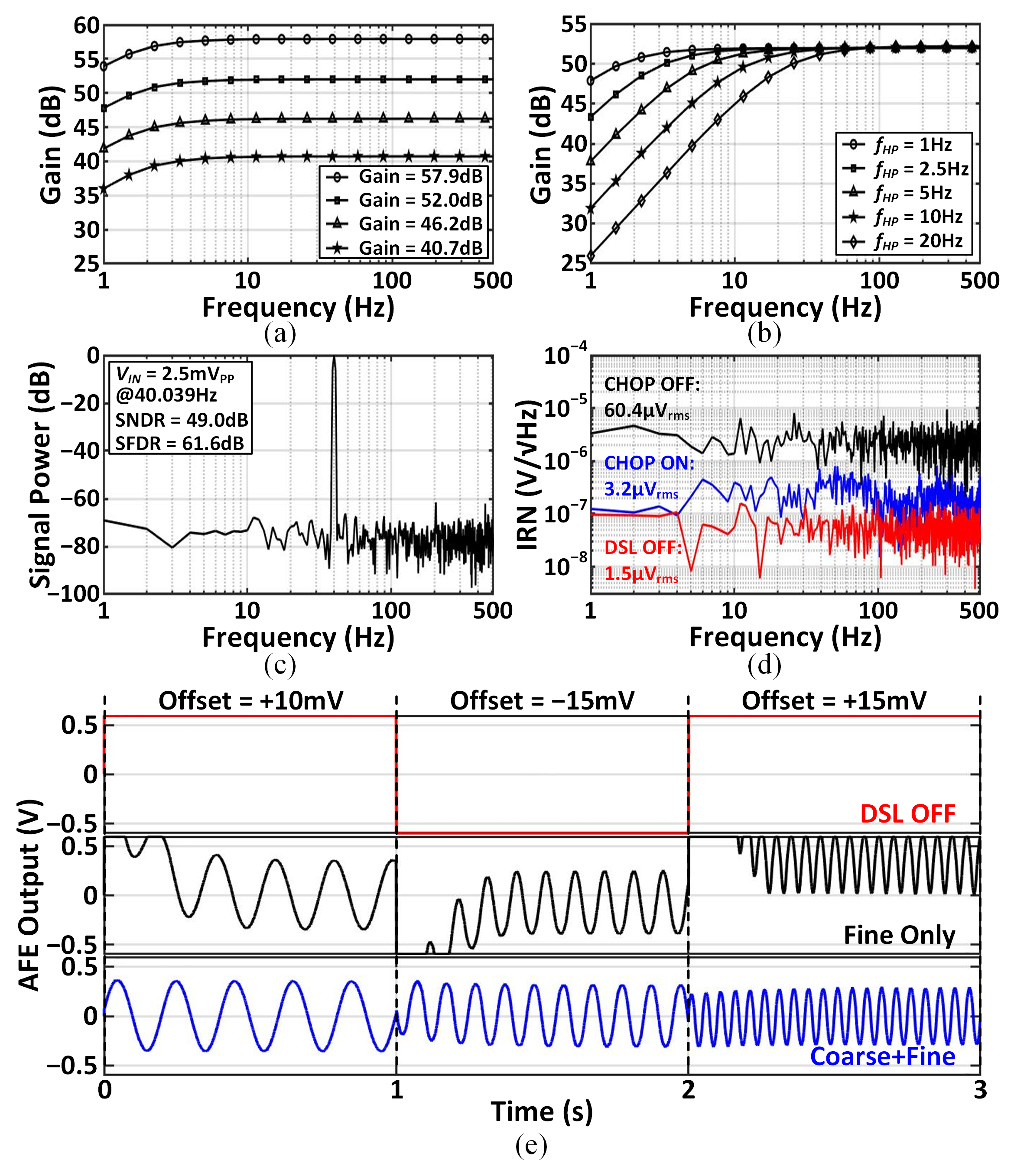}
  \vspace{-6mm}  
  \caption{Measured AFE performance: (a) programmable gain, (b) programmable high-pass pole, (c) output power spectral density (PSD) with a single-tone input, (d) input-referred noise PSD, and (e) two-step fast-settling DSL operation in the presence of abrupt offset changes over three successive 1s windows.}\vspace{0mm}
  \label{f16_AFE_measured}
\end{figure}

\section{Measurement Results}
The SoC was fabricated in a TSMC 65nm 1P9M low-power CMOS process with a chip dimension of 4mm$\times$2mm. The chip micrograph is shown in Fig. \ref{f14_chip}. The 256-channel SoC only occupies an active area of 3.48mm$^2$ (0.014mm$^2$/channel).

\subsection{AFE Characterization}
The 256-channel TDM AFE including the digital DSL occupies an area of 1.1mm$^2$ (0.004mm$^2$/channel) and consumes 387$\upmu$W (1.51$\upmu$W/channel) in training mode. During channel-selective inference, the LNAs and DSLs in the three auxiliary AFE modules are disabled, reducing the AFE power to 182$\upmu$W. Fig. \ref{f16_AFE_measured}(a) shows the measured AFE gain that is programmable between 40.7 and 57.9dB. The location of the high-pass pole can be adjusted by bit-shifting the output of the digital integrator, as demonstrated in Fig. \ref{f16_AFE_measured}(b). With a 2.5mV\textsubscript{PP}, 40.039Hz sine input, the full AFE signal chain including the ADC achieved a signal-to-noise and distortion ratio (SNDR) of 49dB and a spurious-free dynamic range (SFDR) of 61.6dB, measured with 2048-point FFT. The input-referred noise (IRN) performance is presented in Fig. \ref{f16_AFE_measured}(d). Without chopping, the measured IRN was 60.4$\upmu$V\textsubscript{rms} in the 1\text{--}500Hz band. The dominant noise source is the \textit{kT}/\textit{C} noise due to the reset operation for crosstalk reduction, which is in agreement with simulations. When chopping was enabled, the IRN reduced to 3.2$\upmu$V\textsubscript{rms} with the \textit{kT}/\textit{C} and 1/\textit{f} noise up-modulated by the chopper and filtered out by the $\textit{G}_{\textit{m}}$-$\textit{C}$ integrator, as discussed in Section III-C. With the fine DSL disabled, the IRN was measured at 1.5$\upmu$V\textsubscript{rms}. This indicates that when the fine DSL is active, $\sim$2$\times$ noise folding occurs due to the LNA's insufficient bandwidth with respect to the {$\Delta$}{$\Sigma$} frequency \cite{muller2014minimally}. The noise performance could be improved by using a wider LNA bandwidth and lower OSR with a higher-order {$\Delta$}{$\Sigma$} modulator \cite{fathy2021digitally}. Fig. \ref{f16_AFE_measured}(e) demonstrates the fast-settling behavior of the proposed coarse-fine DSL in the presence of abrupt EDO changes over 1s windows, which enables channel-selective inference. With the DSL disabled, the AFE was completely saturated by the offsets. When only the fine DSL was activated, the AFE failed to capture a significant portion of signals during its settling, as other existing DSLs would behave. With a 128kHz chopping frequency, the input impedance was measured to be 24.5M$\Omega$ at 100Hz. The common-mode rejection ratio (CMRR) and power-supply rejection ratio (PSRR) at 50Hz were 70dB and 71dB, respectively. Crosstalk between channels was less than \text{--}79.8dB inter-module and \text{--}72.8dB intra-module. 

\begin{figure}[t]
  \centering
  \includegraphics[width=0.9\columnwidth]{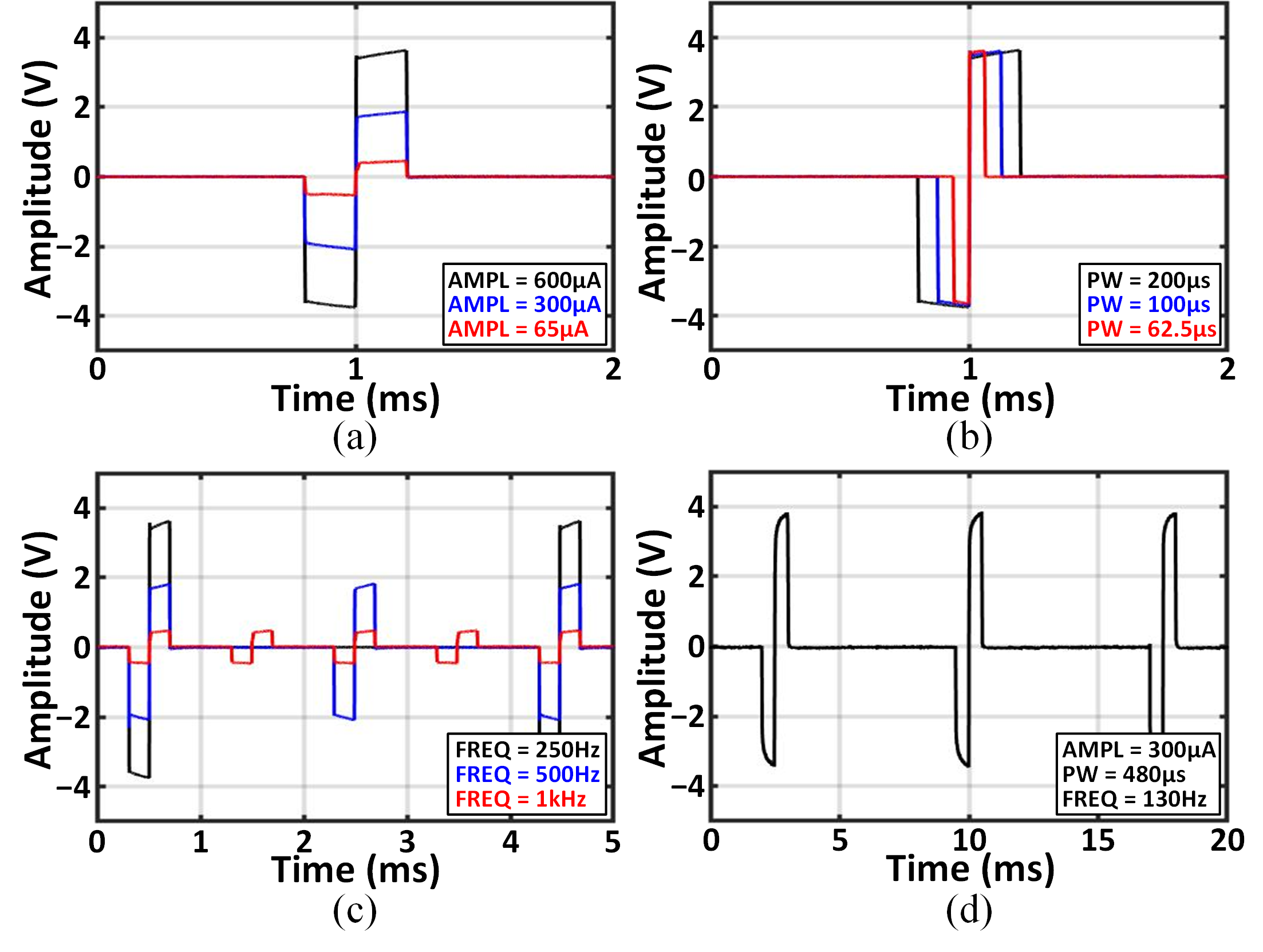}
  \vspace{-3mm}  
  \caption{Biphasic outputs of the neurostimulator:  (a) programmable current amplitude, (b) pulse width, (c) frequency, and (d) \emph{in-vitro} measured output.}\vspace{-3mm}
  \label{f17_STIM_measured}
\end{figure}

\begin{figure*}[t]
  \centering
  \includegraphics[width=2\columnwidth]{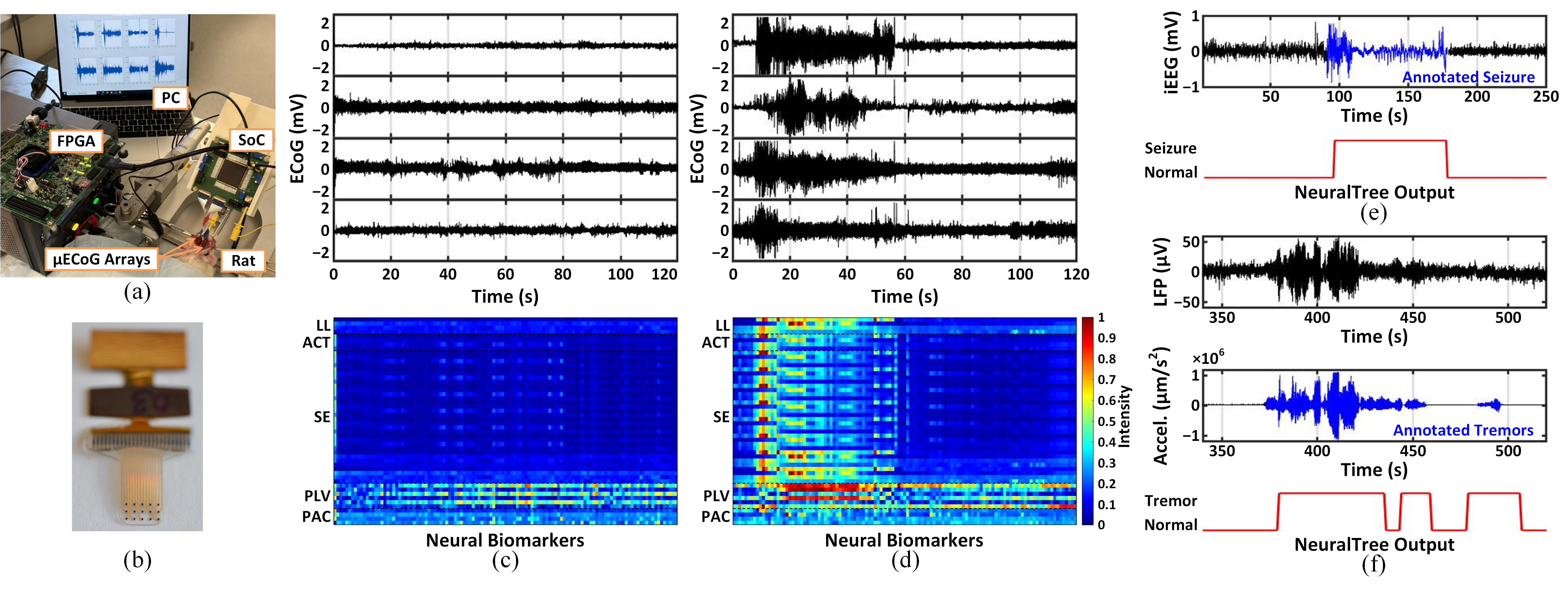}
  \vspace{-3mm}  
  \caption{The SoC validation on a rat model of epilepsy and human datasets: (a) experimental setup for \emph{in-vivo} testing, (b) the 15-channel soft $\upmu$ECoG array with a flexible cable, (c) ECoG recordings and neural biomarker extraction in  a normal state,  (d) ECoG recordings and neural biomarker extraction in a seizure state, (e)  epileptic seizure detection from iEEG, and (f) Parkinsonian tremor detection from LFP.}\vspace{-0mm}
  \label{f17_invivo_human}
\end{figure*}

\subsection{Stimulator Characterization}
The stimulator occupies a small area of 0.05mm$^2$/channel, which is 4$\times$ and 7$\times$ more compact than the stacked architectures in \cite{cheng2018fully} and  \cite{uehlin2020single}, respectively. Figs. \ref{f17_STIM_measured}(a)\text{--}(c) present biphasic stimulation current outputs with different parameters, measured using a 6k$\Omega$+330nF load. Each channel can generate moderate currents ranging from 65$\upmu$A to 600$\upmu$A. With a 640kHz input clock, the  pulse width and frequency are programmable within 9.375$\upmu$s\text{--}203.125$\upmu$s and 9.6Hz\text{--}65kHz, respectively. The stimulator output measured \emph{in vitro} is presented in Fig. \ref{f17_STIM_measured}(d). Under the maximum load current condition, the charge mismatch between the anodic and cathodic phases was measured to be $<$0.1$\%$.

\subsection{\textit{In-vivo} Measurements}
The neural recording and biomarker extraction capabilities of the SoC were validated \emph{in vivo} in the experimental setup shown in Fig. \ref{f17_invivo_human}(a). The animal experiments were performed with the approval of all experimental and ethical protocols and regulations granted by the Veterinary Office of the Canton of Geneva, Switzerland, under License No. GE/33A (33223). All procedures were in accordance with the Regulations of the Animal Welfare Act (SR 455) and Animal Welfare Ordinance (SR 455.1). We implanted two 15-channel 200$\upmu$m-diameter soft $\upmu$ECoG arrays, shown in Fig. \ref{f17_invivo_human}(b), into the somatosensory cortex of a Lewis rat. The $\upmu$ECoG arrays were fabricated using an e-dura technology with gold thin films~\cite{minev2015electronic}. Pentylenetetrazol (20mg) was injected intraperitoneally to the anesthetized rat to induce seizures \cite{velisek1992pentylenetetrazol}. Figs. \ref{f17_invivo_human}(c) and (d) present ECoG recordings and neural biomarkers during a normal and a seizure state, respectively. Prominent increases in temporal and spectral biomarkers at the seizure onset and strong cross-channel phase synchronization were observed during a seizure event in a short acute recording session. In future work, the real-time seizure detection capability of the SoC will be further validated \emph{in vivo} with a collection of sufficient seizure activity and objective seizure annotation. 

\subsection{Epileptic Seizure Detection}
The classification performance of the SoC was validated on the CHB-MIT EEG \cite{goldberger2000physiobank} and iEEG.org \cite{ieeg} datasets of epilepsy patients. We analyzed 983-hour EEG recordings of 24 patients and 596-hour iEEG recordings of 6 patients, which contain 176 and 49 annotated seizures, respectively. Blockwise data partitioning  was used to avoid data leakage from training to inference \cite{shoaran2018energy}. We performed 5-fold cross-validation for most patients and adopted a leave-one-out approach for patients with fewer than 5 seizures. The number of correctly detected seizures was counted to assess the sensitivity, while the specificity was calculated based on the window-based true negative rate averaged over multiple runs.

In training mode, each patient's multi-channel  neural data (18\text{--}28 EEG and 47\text{--}108 iEEG channels) were fed to the AFE. The digitized AFE outputs were processed offline to extract features using a bit-accurate FEE model in MATLAB for training the classifier. The trained NeuralTree parameters were then stored to the on-chip memory for inference, and the NeuralTree performance on the test data was evaluated. The SoC achieved 95.6$\%$/94$\%$ sensitivity and 96.8$\%$/96.9$\%$ specificity on the EEG and iEEG datasets, respectively.
The SoC's seizure detection performance on an epileptic patient is demonstrated in Fig. \ref{f17_invivo_human}(e). 

\begin{table*}[t]
  \centering  
  \vspace{-3mm}    
  \caption{Comparison with the State-of-the-Art Neural Interface SoCs with On-Chip ML}
  \includegraphics[width=2\columnwidth]{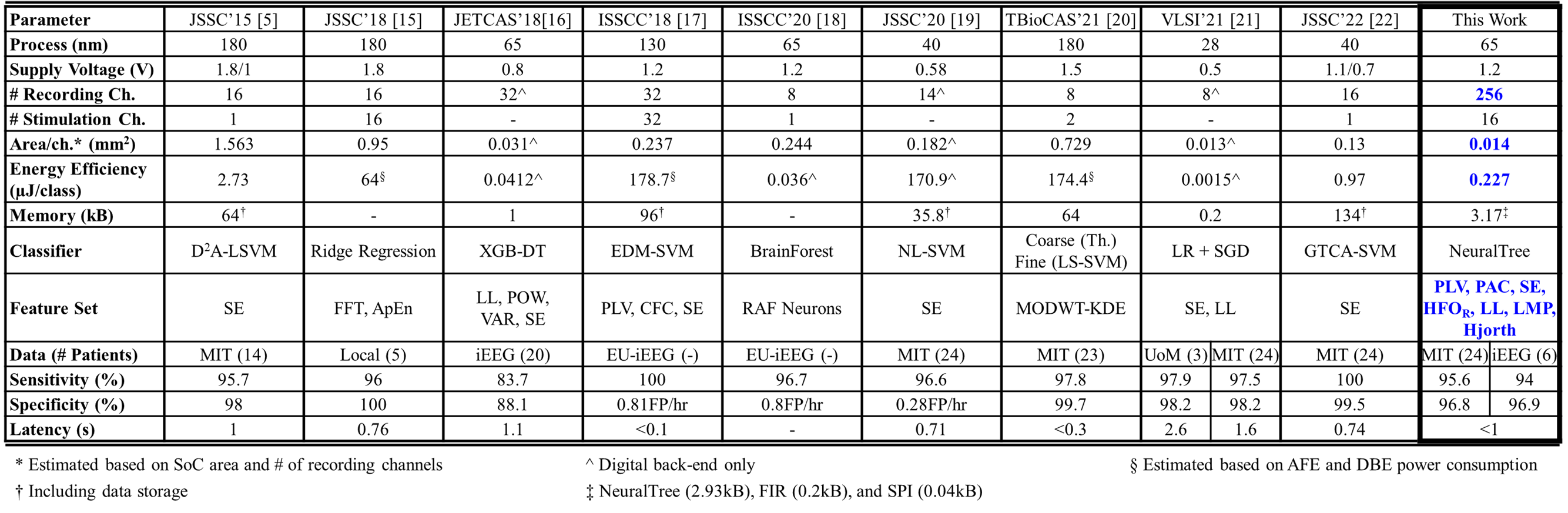}\vspace{-3mm}  
  \label{t2_comparison}
\end{table*}

\subsection{Parkinsonian Tremor Detection}
The SoC's performance was further validated on a PD patient with rest-state tremor recruited by the University of Oxford \cite{yao2020improved}. A 4-channel DBS lead was implanted into the subthalamic nucleus to collect LFPs, while the acceleration of the contralateral limb was used to label the  tremor. 
Window-based true positive and negative rates were used to assess the sensitivity and specificity, respectively, using a 5-fold cross-validation. 
The SoC achieved 82.6\% sensitivity and 78.4\% specificity. Fig. \ref{f17_invivo_human}(f) presents an example of the SoC's tremor detection performance, where tremor states with inconspicuous neural activity were successfully detected by the NeuralTree. To the best of our knowledge, this is the first demonstration of PD tremor detection with an on-chip classifier. 

\subsection{Comparison with the State-of-the-Art}
Table. \ref{t2_comparison} compares the proposed SoC with the state-of-the-art seizure detection SoCs with on-chip ML. The 256-channel SoC achieves an 8$\times$ improvement in channel count, 9.3$\times$ in per-channel area, and 4.3$\times$ in system energy efficiency  over the state-of-the-art. The compressed NeuralTree  takes up 2.93kB out of the 3.17kB  on-chip memory, enabling more efficient memory utilization than  SVM classifiers \cite{altaf201516, o2018recursive, huang20191, wang2021closed, zhang2022patient}. For instance, the recent SVM classifier in \cite{zhang2022patient} utilized 70kB of memory to store 256 support vectors for 16 channels. The proposed SoC achieves  better multi-channel scalability than previous DT-based SoCs \cite{shoaran2018energy, o202026} thanks to the end-to-end TDM  implementation. Moreover, the SoC integrates the broadest range of neural biomarkers reported so far to provide greater flexibility for multiple neural classification tasks. 

\section{Conclusion}
To pave the way towards next-generation closed-loop neural interfaces, this article presented a highly scalable, versatile neural activity classification and closed-loop neuromodulation SoC integrating an area-efficient dynamically addressable 256-channel mixed-signal front-end, multi-symptom biomarker extraction, an energy-aware NeuralTree classifier, and a 16-channel HV compliant neurostimulator. A channel-selective inference scheme was introduced to overcome the limited  scalability and low hardware efficiency of the existing SoCs. Through aggressive system-level time-division multiplexing and energy-efficient circuit-algorithm co-design, the proposed 256-channel SoC achieved the highest level of integration reported so far, as well as the highest energy and area efficiency. The versatility of the SoC was demonstrated using human epilepsy and Parkinson's datasets with multiple signal modalities (EEG, iEEG, and LFP). This high-channel-count SoC with the multi-class NeuralTree model can be further used for motor intention decoding in prosthetic BMIs.

\section*{Acknowledgment}
The authors acknowledge Ivan Furfaro for his contribution to the \emph{in-vivo} experimental setup. This work was partially supported by the National Institute of Mental Health Grant R01-MH-123634 and funding from EPFL.
 
\ifCLASSOPTIONcaptionsoff
  \newpage
\fi

\bibliographystyle{IEEEtran.bst}
\bibliography{cit}

\end{document}